\documentclass[floatfix,%
 reprint,
 amsmath,amssymb,
 aps,
 superscriptaddress,
]{revtex4-2}
\usepackage{lmodern}
\usepackage{mathtools}
\usepackage[retainorgcmds]{IEEEtrantools}
\usepackage{graphicx}
\usepackage{dcolumn}
\usepackage{bm}
\usepackage{color}
\usepackage{multirow}
\usepackage{booktabs}
\usepackage{array}
\usepackage{amsmath}      
\usepackage{amsfonts}     
\usepackage{amssymb}      
\usepackage{siunitx}     
\usepackage{adjustbox}
\usepackage{mathtools} 
\usepackage[caption=false]{subfig} 
\usepackage{graphicx}
\usepackage{url}
\usepackage{color, soul}

\begin{document}


\title{Accurate neural network emulator for primordial light element abundances}

\author{Fan Zhang}
\affiliation{State Key Laboratory of Ocean Sensing \& Ocean College, Zhejiang University, Zhoushan, Zhejiang 316000, China}
\affiliation{Kavli Institute for Astrophysics and Space Research, Massachusetts Institute of Technology, Cambridge, MA, 02139, USA} 

\author{Hang Diao}
\affiliation{State Key Laboratory of Ocean Sensing \& Ocean College, Zhejiang University, Zhoushan, Zhejiang 316000, China}

\author{Bohua Li}
\email{bohuali@gxu.edu.cn}
\affiliation{Guangxi Key Laboratory for Relativistic Astrophysics, School of Physical Science and Technology, Guangxi University, Nanning, Guangxi, 530004, China}

\author{Joel Meyers}
\affiliation{Department of Physics, Southern Methodist University, Dallas, TX 75275, USA}

\author{Paul R. Shapiro}
\affiliation{Department of Astronomy, The University of Texas at Austin, Austin, TX 78712, USA}


\begin{abstract}
Big-Bang Nucleosynthesis (BBN) predictions of primordial light-element abundances offer a powerful probe of early-Universe physics. However, high-accuracy numerical BBN calculations have become a major computational bottleneck for large-scale cosmological inferences due to the complex nuclear network. Here we present \verb|BBNet|, a fast and accurate deep learning emulator for primordial abundances. The training data are generated by full numerical calculations using two public BBN codes, \verb|PArthENoPE| and \verb|AlterBBN|, modified to accommodate extended cosmologies that include dark radiation and a stiff equation of state. The network employs a residual multi-head architecture to capture convoluted physical relationships. \verb|BBNet| produces primordial helium-4 and deuterium abundances with negligible errors in milliseconds per sample, achieving a speed-up of up to $10^4$ times relative to first-principles solvers while remaining unbiased over wide parameter ranges. Therefore, our emulator can supersede traditional simplified numerical prescriptions that compromise accuracy for speed. Based on extensive assessments of its performance, we conclude that \verb|BBNet| is an optimal solution to the theoretical prediction of primordial element abundances. It will serve as a reliable tool for precision cosmology and new-physics searches.
\end{abstract}


\maketitle

\section{Introduction}
\label{sec:intro}

Primordial abundances of the light elements
produced during big-bang nucleosynthesis (BBN)
present a direct probe of the physical conditions in the early Universe
\cite{1949PhRv...75.1089A,1998RvMP...70..303S,2006NuPhA.777..208F}.
The agreement between their predictions and observations
supports the concordance hot big-bang cosmology
\cite{2007ARNPS..57..463S,2016RvMP...88a5004C,2020JCAP...03..010F}.
In particular, the primordial helium and deuterium abundances
together offer one of the most stringent constraints
on the standard $\Lambda$CDM model (e.g., on the baryon density)
as well as new physics beyond the Standard Model (BSM)
\cite{1995Sci...267..192C,2009PhR...472....1I,2010ARNPS..60..539P,2022JCAP...10..046Y}.

The precision of the BBN constraints depends on
accurate theoretical calculations of the BBN reaction network
\cite{1967ApJ...148....3W,2004JCAP...12..010S,2012ApJ...744..158C}.
However, the nuclear reaction rates that enter the BBN network
are subject to laboratory measurement uncertainties
\cite{2011RvMP...83..195A,2013NuPhA.918...61X,2016ApJ...831..107I,2021ApJ...923...49M},
and the propagated uncertainties on the primordial element abundances
from BBN calculations are often significant
\cite{1998PhRvD..58f3506F,2016PhRvL.116j2501M,2020Natur.587..210M,2021JCAP...03..046Y}.
In the case of the D/H abundance, the theoretical uncertainty
is in fact the dominant source of error
compared with the uncertainty from astrophysical measurements
\cite{2015PhRvD..92l3526C,2018ApJ...855..102C,2021JCAP...04..020P}.
Additional uncertainties may arise from
the prescription of neutrino physics during weak decoupling, 
which partially overlaps the early stage of BBN
\cite{2016PhRvD..93h3522G,2019JCAP...02..007E}.
Therefore, it is crucial that the BBN calculation itself,
which involves solving the set of coupled ordinary differential equations (ODEs)
that describe the BBN network,
does not add to the theoretical errors.

Modern numerical solvers have been developed to meet this need,
including public BBN codes \verb|PArthENoPE|
\cite{2008CoPhC.178..956P,2018CoPhC.233..237C,2022CoPhC.27108205G},
\verb|AlterBBN| \cite{2012CoPhC.183.1822A,2018arXiv180611095A},
\verb|PRIMAT| \cite{2018PhR...754....1P}, \verb|PRyMordial| \cite{2024EPJC...84...86B},
\verb|pyBBN| \cite{2020JCAP...11..056S}, etc.
Most of these methods are updates of the earlier Kawano code,
\verb|NUC123| \cite{1992STIN...9225163K},
which is itself based on the historical Wagoner code \cite{1969ApJS...18..247W,1973ApJ...179..343W}.
They yield accurate solutions of primordial abundances,
while some of the codes entertain various BSM scenarios.
However, the computational demands of these first-principles methods are high;
even the fastest BBN codes should take $\mathcal{O}(0.1)$ seconds
for a single execution in simplified settings and longer in full settings.
Taking \verb|PArthENoPE| for example, a single run can cost tens of seconds on a CPU
with its complete 26-nuclide configuration.
Such costs pose a serious challenge to parameter inference tasks
based on standard techniques such as Markov chain Monte Carlo (MCMC) sampling,
which typically require thousands of evaluations to explore posterior distributions.
The burden is exacerbated in BSM scenarios that involve additional parameters,
e.g., variations of fundamental constants, dark radiation,
nonzero neutrino chemical potentials, sterile neutrinos,
\cite{2005PhRvD..71l3524B,2015PhRvD..91h3505N,2018arXiv180611095A,
2020JCAP...11..056S,2022CoPhC.27108205G,2024EPJC...84...86B}, etc.

This computational bottleneck often forces
compromises in the numerical treatments of BBN calculations.
Some methods adopt or provide simplified nuclear reaction networks,
achieving faster computation at the cost of introducing systematic biases.
For example, even the recently developed codes \verb|PRyMordial| and \verb|LINX|
consider only 62 reactions in their ``full'' network with 12 nuclides
\cite{2024arXiv240814538G}, in contrast to the ``complete'' network of 100 reactions
in \verb|PArthENoPE| and \verb|AlterBBN|.
Another approach to speed-up is
to use less accurate integration algorithms for the ODEs.
\verb|AlterBBN| offers a spectrum of ODE solvers
to explore the speed-accuracy trade-off \cite{2018arXiv180611095A}.
Apart from these efforts, the interface between the BBN code
and Bayesian inference applications are optimized in the \verb|LINX| package,
which enables gradient-assisted inference
that achieves faster convergence than MCMC \cite{2024arXiv240814538G}.

In this paper, we introduce \verb|BBNet|, a deep learning emulator
for efficient and accurate BBN computations of primordial element abundances.
We employ multi-layer perceptrons with residual connections
as the main architecture of \verb|BBNet|,
together with a multi-head attention module
\cite{He2016, Goodfellow-et-al-2016,Vaswani2017}.
Based on the training data generated by existing BBN codes,
\verb|PArthENoPE| and \verb|AlterBBN|,
our emulator can output the primordial helium and deuterium abundances
in $\sim 6$\,ms per sample on an NVIDIA A100 GPU
and in $\sim 2.5$\,ms per sample with batch processing.
This corresponds to a speed-up of $\mathcal{O}(10^4)$ times
compared with first-principles BBN calculations.

The high computational efficiency of \verb|BBNet|
is particularly helpful in exploring scenarios that involve BSM physics.
In its current version, we consider an extra dark radiation component,
parameterized by the number of extra relativistic species,
and a possible kination/stiff phase in the early Universe \cite{2010PhRvD..82h3501D}.
The latter is a pre-BBN era in the expansion history
during which the total energy density of the Universe
is dominated by the kinetic energy of a scalar field
\cite{1985PhLB..155..232B,1993PhLB..315...40S,1997PhRvD..55.1875J,2025PhRvL.135j1002E}.
During kination or a stiff phase, the equation of state (EoS)
of the Universe is given by that of a stiff fluid,
$\bar{w} \equiv P / \bar{\rho} = 1$.
It can occur in extended cosmological models
with quintessence, ultralight scalar field dark matter, QCD axion
\cite{1999PhRvD..59f3505P,2014PhRvD..89h3536L,2025SCPMA..6880409Y,2020PhRvL.124y1802C}, etc.

\verb|BBNet| attains exceptional emulation accuracy
in addition to its speed advantage.
The relative root mean square (RMS) errors
between the predicted primordial abundances, $Y_\mathrm{P}$ and D/H,
and their ground-truth values from the BBN calculations
are typically of the order of 0.1\% or less.
The predictions of the emulator are also unbiased,
achieving a mean percentage error well below 0.1\%.
The errors from \verb|BBNet| emulations are thus negligible compared with
current theoretical and observational uncertainties of $Y_\mathrm{P}$ and D/H.
The unprecedented speed and accuracy of \verb|BBNet|
present a solution to the BBN computational bottleneck,
reducing the cost of high-dimensional cosmological analyses by orders of magnitude.
By integrating \verb|BBNet| into MCMC pipelines,
one can perform efficient joint analyses of large cosmological data sets
based on extended models without compromising the accuracy of BBN computations.

These capabilities of \verb|BBNet| are relevant for both current and future analyses. 
As we will see, the systematic biases induced by widely-used simplified BBN solvers
are no less than the present observational uncertainties for $Y_\mathrm{P}$ and D/H 
from the 2024 report of the Particle Data Group (PDG 2024) \cite{Fields:2025}.
By contrast, the errors of \verb|BBNet| emulations are well below
those uncertainties by more than an order of magnitude.
In the future, the ANDES spectrograph on the 39\,m Extremely Large Telescope (ELT) is expected to determine D/H 
to $0.1\,\%-0.2\,\%$ precision \cite{2024ExA....57....5M,2024SPIE13096E..13M}. 
The exceptional accuracy of \verb|BBNet| (with sub-per-mille RMS relative errors)
is therefore crucial to keep the computational errors safely below the observational uncertainties.

The remainder of this paper is organized as follows.
Sec.~\ref{sec:bbn_physics} presents a brief overview of BBN physics
and then describes the BBN codes and the cosmological model considered in this work.
Sec.~\ref{sec:methodology} provides a detailed description of our data generation strategy and the architecture of the \verb|BBNet| neural network emulator.
Sec.~\ref{sec:training} discusses the optimization protocols,
regularization techniques, and validation procedures for training the emulator.
In Sec.~\ref{sec:results}, we present comprehensive
performance benchmarks and accuracy assessments.
Sec.~\ref{sec:comparison} compares the performance of \verb|BBNet|
with those of existing fast approximation modes
internal to \verb|PArthENoPE| and \verb|AlterBBN|.
Finally, Sec.~\ref{sec:conclusion} summarizes our key conclusions
and proposes directions for future research.

\section{BBN calculation and cosmological model}
\label{sec:bbn_physics}

This section reviews the key aspects of BBN,
including the thermal history of the early Universe, the nuclear reaction network,
and the numerical techniques required by BBN calculations.
We then describe the BSM physics that we take into account in this work.
We correspondingly modify the public BBN codes, \verb|PArthENoPE| and \verb|AlterBBN|,
to implement the extended cosmological model.

\subsection{BBN as the primordial nuclear reactor}

The Universe had cooled to temperatures below $10^{10}\,\si{\kelvin}$
in about 1\,s after the Big Bang \cite{2008cosm.book.....W}.
At this epoch, the rate of weak interactions responsible for neutron–proton conversion
started to fall below the Hubble expansion rate,
leading to the freeze-out of the neutron-to-proton ratio, $n/p$
\cite{2002PhR...370..333D}.
After freeze-out, beta decays of free neutrons
slightly decreased the ratio to $n/p\approx 1/7$,
setting the stage for nucleosynthesis \cite{2016RvMP...88a5004C}.

The first few reactions in the nuclear chain are concerned with
the synthesis of deuterium via $p + n \leftrightarrow d + \gamma$,
which had to overcome the so-called deuterium bottleneck
due to the huge number of photons at $T\gtrsim 0.1$\,MeV.
Only at lower temperatures can deuterium be massively produced,
and a rapid sequence of BBN reactions then commenced.
These reactions efficiently converted nearly all available neutrons
into helium-4 ($^4\mathrm{He}$) within three minutes after the Big Bang,
since it is the most tightly bound nuclide among light elements
\cite{2006IJMPE..15....1S}.
The primordial helium abundance expressed
in terms of the (approximate) mass fraction,
$Y_\mathrm{P}\equiv 4\,n_\mathrm{^4He}/n_\mathrm{b}$,
is thus primarily determined by the neutron-to-proton ratio at freeze-out;
$Y_\mathrm{P}\simeq 0.25$ in standard BBN.
Here $n_\mathrm{b}$ is the total number density of baryons.
Other light nuclides are synthesized at the same time,
including helium-3 and lithium-7.

In standard BBN, the final values of the light element abundances
are sensitive to two physical parameters:
(i) the baryon-to-photon ratio, $\eta \equiv n_\mathrm{b}/n_\gamma$,
and (ii) the free neutron lifetime, $\tau_n$.
The baryon-to-photon ratio describes the baryon density of the Universe.
A higher value of $\eta$ leads to more efficient deuterium burning,
resulting in a lower primordial deuterium abundance,
$\mathrm{D/H}\equiv n_\mathrm{D}/n_\mathrm{H}$
\cite{2007ARNPS..57..463S,1995Sci...267..192C}.
Meanwhile, the longer the neutron lifetime,
the higher the value of $n/p$ due to fewer decays
and hence higher $Y_\mathrm{P}$.
Therefore, even though these two parameters are well measured
by other astronomical observations or laboratory experiments
\cite{2020A&A...641A...6P,2021PhRvL.127p2501G},
BBN provides an independent verification.

\subsection{Nuclear network and differential equations}

Modern BBN calculations are computationally intensive
due to the need to evolve an extensive nuclear reaction network,
often composed of tens of nuclear species
linked by over a hundred reactions \cite{1967ApJ...148....3W}.
For example, \verb|PArthENoPE| and \verb|AlterBBN|
consider the same 100 reactions for 26 nuclides
\cite{2008CoPhC.178..956P,2012CoPhC.183.1822A},
while \verb|PRIMAT| evolves as many as 391 reactions for 59 nuclides
\cite{2012ApJ...744..158C,2018PhR...754....1P}.
In each BBN code, the corresponding system of coupled ODEs
is integrated to obtain the primordial element abundances.
Defining $Y_i\equiv n_i/n_\mathrm{b}$ for the $i$th nuclear species
($n_i$ is its number density),
one can express the general form of the ODE
that describes a reaction in the network as
\begin{equation}\label{eq:nuc}
\begin{split}
\dot{Y}_{i_1} = &
\sum_{i_2 \ldots i_p, j_1 \ldots j_q} 
N_{i_1}\,\Bigg( 
\Gamma_{j_1 \ldots j_q \rightarrow i_1 \ldots i_p} 
\frac{Y_{j_1}^{N_{j_1}} \cdots Y_{j_q}^{N_{j_q}}}{N_{j_1}! \cdots N_{j_q}!} \\
& \qquad -\Gamma_{i_1 \ldots i_p \rightarrow j_1 \ldots j_q} 
\frac{Y_{i_1}^{N_{i_1}} \cdots Y_{i_p}^{N_{i_p}}}{N_{i_1}! \cdots N_{i_p}!}
\Bigg)\,,
\end{split}
\end{equation}
without restricting to two-body reactions
\cite{1969ApJS...18..247W,2018PhR...754....1P}.
Here $N_i$ is the stoichiometric coefficient of species $i$
(the number of the $i$th nuclide) in the reaction
and $\Gamma_{i_1 \ldots i_p \rightarrow j_1 \ldots j_q}$
denotes the nuclear reaction rates,
proportional to the thermally averaged cross sections, $\langle\sigma v\rangle$.
Eq.~(\ref{eq:nuc}) results from the Boltzmann equation.
Nontrivial phases in the thermal history of the early Universe,
such as electron-positron annihilation and neutrino decoupling,
affect the nuclear network via the energy densities of relevant species
and the Hubble expansion rate \cite{2005NuPhB.729..221M,2016PhRvD..93h3522G}.

The resulting system of ODEs is highly stiff
because the timescales of nuclear reactions
span many orders of magnitude \cite{2024EPJC...84...86B}.
Its numerical integration requires extremely small step sizes
when rapid burning occurs, which is otherwise suboptimal
for accurately resolving slower weak interactions
at the onset of BBN calculations \cite{2018PhR...754....1P,2016RvMP...88a5004C}.
In the meantime, stringent tolerance is needed
to avoid secular drifts in the predicted abundances.
Furthermore, each integration step often involves
repeated interpolations of a grid of reaction rates
and an implicit solution to the temperature-entropy evolution
\cite{2018PhR...754....1P,2016RvMP...88a5004C,2016PhRvD..93h3522G}.
Hence, all of the above factors and the sheer dimensionality of the network
combine to make traditional first-principles BBN computations
exceedingly time-consuming \cite{2018CoPhC.233..237C}.
This motivates our deep learning emulator approach.

\subsection{BBN codes and cosmological model}
\label{sec:bbn_codes}

In this work, we use the public BBN codes
\verb|PArthENoPE| v3.0 and \verb|AlterBBN| v2.2
to generate training data for our emulator
\cite{2022CoPhC.27108205G,2018arXiv180611095A}.
To ensure a consistent implementation of nuclear physics,
both codes are modified to use identical reaction rates for the following
key reactions that strongly impact $Y_\mathrm{P}$ and $\mathrm{D/H}$:
\begin{align}
p + n &\leftrightarrow \gamma + {}^{2}\mathrm{H} & (\mathrm{PNG}) \notag \\
{}^{2}\mathrm{H} + p &\leftrightarrow \gamma + {}^{3}\mathrm{He} & (\mathrm{DPG}) \notag \\
{}^{2}\mathrm{H} + {}^{2}\mathrm{H} &\leftrightarrow n + {}^{3}\mathrm{He} & (\mathrm{DDN}) \notag \\
{}^{2}\mathrm{H} + {}^{2}\mathrm{H} &\leftrightarrow p + {}^{3}\mathrm{H} & (\mathrm{DDP}) \notag
\end{align}
These processes dominate the synthesis of deuterium and ${}^4$He during BBN,
contributing to most of the theoretical uncertainties on D/H and $Y_\mathrm{P}$
\cite{2004JCAP...12..010S,2021JCAP...03..046Y}.
We consider the theoretical PNG rate calculated in Ref.~\cite{2006PhRvC..74b5809A},
which is adopted by \verb|AlterBBN|.
For the other three reaction rates, we adopt the high-precision fits
used by \verb|PArthENoPE| based on the compilation of experimental data
in Ref.~\cite{2021JCAP...04..020P},
which includes the recent LUNA results on the DPG rate \cite{2020Natur.587..210M}.

The resulting primordial element abundances from BBN calculations
depend on the cosmological model;
they can be sensitive to parameters
of both the standard $\Lambda$CDM model and extended scenarios.
Regarding the former, both \verb|PArthENoPE| and \verb|AlterBBN| take as input
the baryon-to-photon ratio, $\eta$, or $\eta_{10}\equiv\eta\times10^{10}$.
It is equivalent to the cosmic baryon density, $\Omega_\mathrm{b}h^2$,
according to \cite{2016RvMP...88a5004C}
\begin{equation}\label{eq:eta_definition}
    \eta_{10} = 273.3036\,\Omega_{\mathrm{b}}h^2
    \left(1 + 7.16958\times10^{-3}\,Y_\mathrm{P}\right).
\end{equation}
We apply this relationship to convert between $\Omega_\mathrm{b}h^2$
and $\eta_{10}$ (or $\eta$) in our modifications of both BBN codes.

For BSM scenarios, the extra relativistic degrees of freedom
are parameterized by $\Delta N_\mathrm{eff}$ as usual, 
while the standard effective number of relativistic species
is $N^\mathrm{SM}_\mathrm{eff}=3.044$,
accounting for the three neutrinos in the Standard Model (SM)
\cite{2021JCAP...04..073B,2019JCAP...07..014G,2020JCAP...08..012A}.
In addition, we consider a possible pre-BBN kination/stiff phase.
The energy density of a stiff fluid redshifts as $a^{-6}$
so that the universe transitions from the stiff phase
to the radiation-dominated (RD) era at early times.
Following Refs.~\cite{2010PhRvD..82h3501D,2025ApJ...985..117L},
this stiff-to-radiation transition is parameterized
by the ratio of the stiff-fluid energy density to the photon energy density
at $T_\mathrm{10}\equiv 10$\,MeV,
\begin{equation}
    \kappa_{10} \equiv \left(\frac{\rho_\mathrm{s}}{\rho_\gamma}\right)_{T=10\,\mathrm{MeV}}\,.
\label{eq:kappa10_definition}
\end{equation}
The impacts of a stiff phase on the primordial element abundances
are previously examined by some of us
in Refs.~\cite{2021JCAP...10..024L,2017PhRvD..96f3505L,2014PhRvD..89h3536L}.

\begin{figure*}[t]
  \centering
  \subfloat[\label{fig:compare_al_pe_kappa}]{
    \includegraphics[width=0.45\textwidth,height=0.6\textwidth,keepaspectratio=false,trim=4 4 4 4,clip]{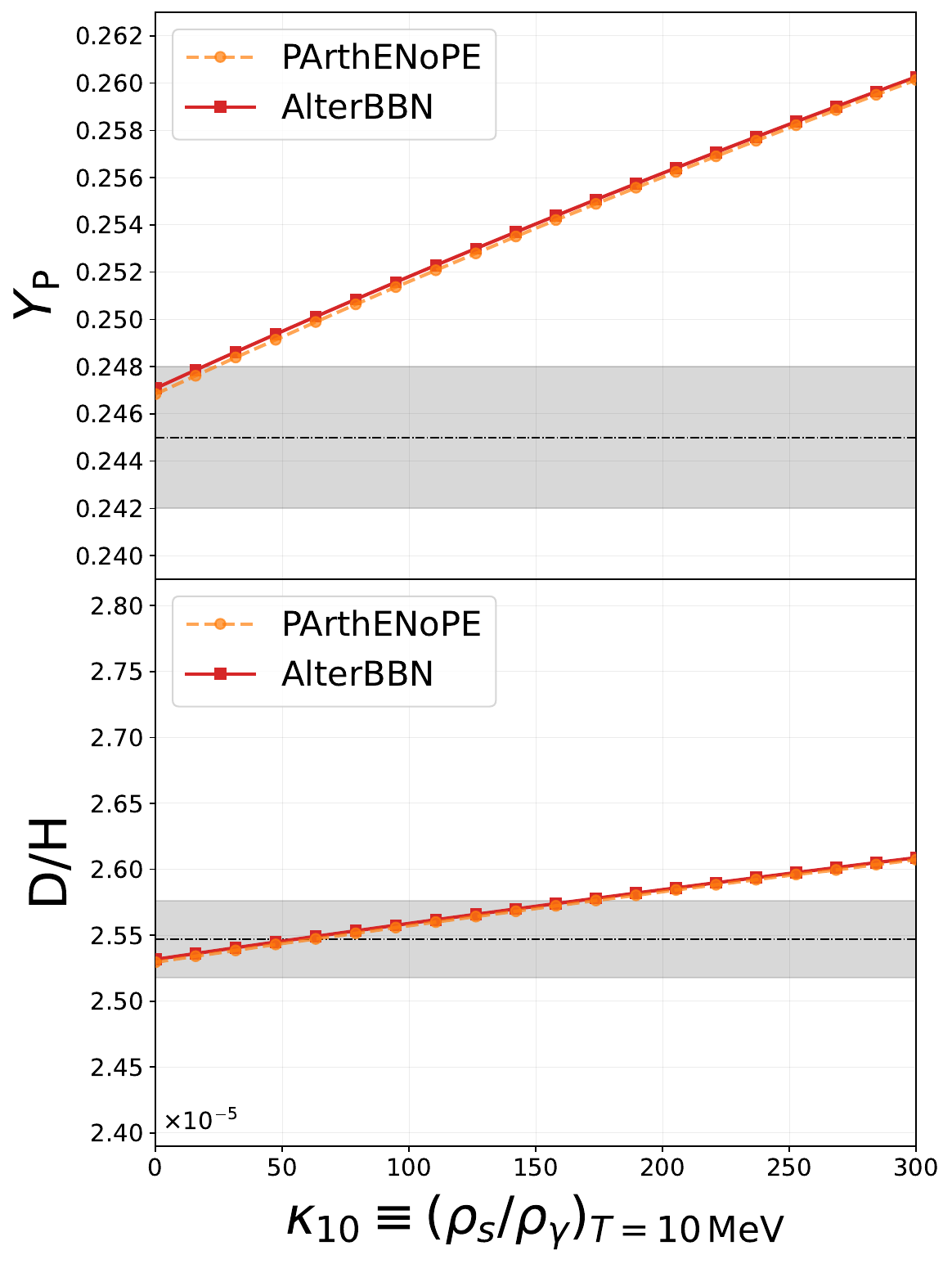}
  }\hspace{0.05\columnwidth}
  \subfloat[\label{fig:compare_al_pe_dnnu}]{    
    \includegraphics[width=0.45\textwidth,height=0.6\textwidth,keepaspectratio=false,trim=4 4 4 4,clip]{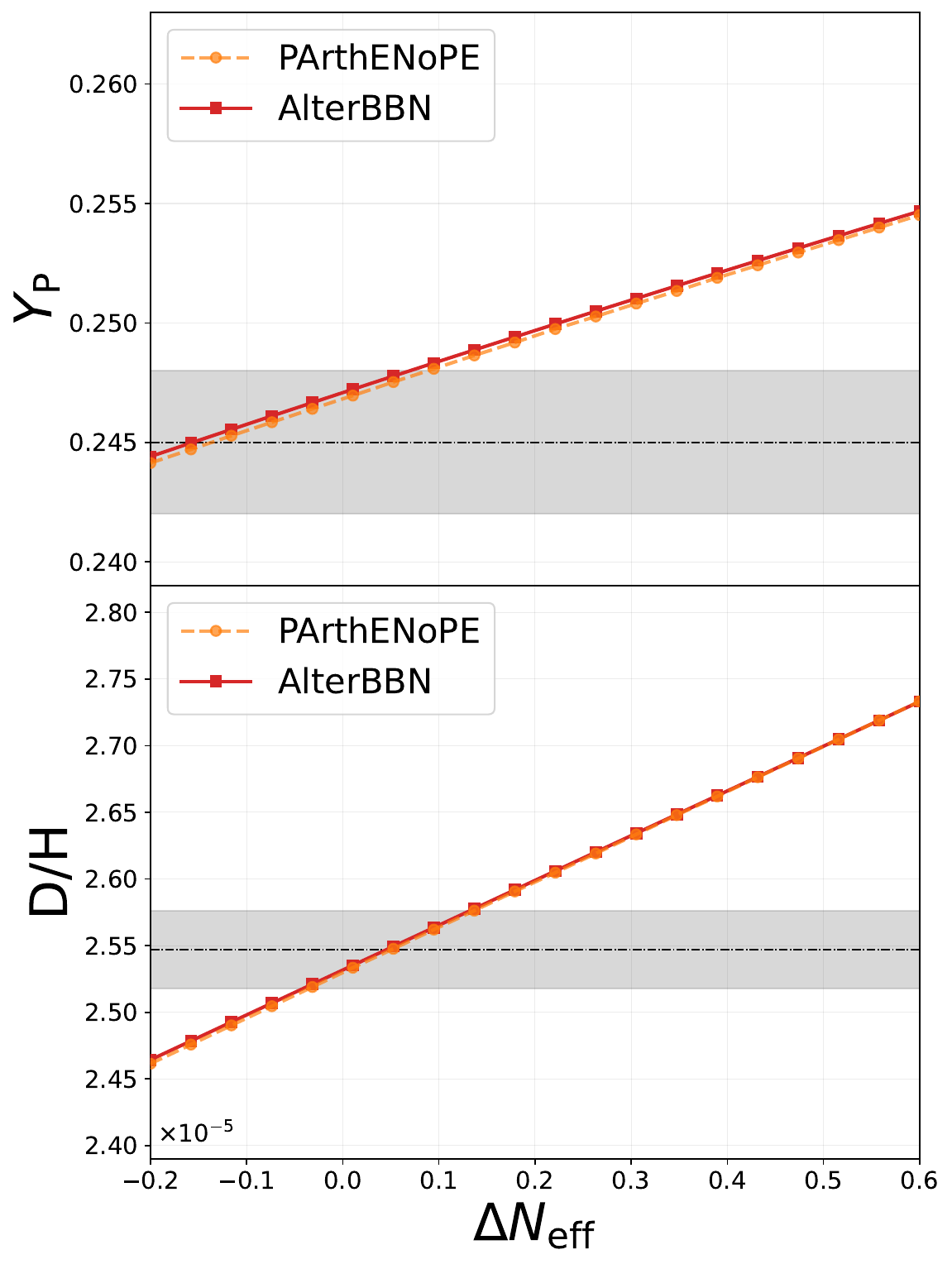}
  }
  \caption{Consensus of primordial element abundances
  computed by our modified PArthENoPE (orange dashed lines with circle markers)
  and AlterBBN (red solid lines with square markers).
    Each panel contains two stacked subplots
    for $Y_\mathrm{P}$ and $\mathrm{D/H}$, respectively.
    Fig.~\ref{fig:compare_al_pe_kappa} displays their relations
    with the density parameter of the stiff fluid,
    $\kappa_{10} \!\equiv\! (\rho_\mathrm{s}/\rho_\gamma)_{T=10\,\mathrm{MeV}}$, while Fig.~\ref{fig:compare_al_pe_dnnu} shows the relations
    with the effective number of extra relativistic species, $\Delta N_\mathrm{eff}$.
    The gray shaded bands denote the $1\sigma$ observational constraints
    from PDG 2024 \cite{Fields:2025},
    and the black dashed lines mark the corresponding central values.}
  \label{fig:bbn_code_comparison}
\end{figure*}

In this work, we use $\Delta N_\mathrm{eff}$ and $\kappa_{10}$ as free parameters.
Our modified \verb|PArthENoPE| takes them as input parameters directly.
On the other hand, to adapt to the code structure of \verb|AlterBBN|,
our modified version of \verb|AlterBBN| describes
the extra radiation and the stiff phase
by equivalent input parameters $\kappa_{\mathrm{rad},\,i}$ and $\kappa_{\mathrm{s},\,i}$,
respectively. They are defined as
\begin{align}
    \kappa_{\mathrm{rad},\,i} & \equiv \Delta N_\mathrm{eff}\cdot\frac{7}{8} \left(\frac{4}{11}\right)^{4/3} \left(\frac {T_\mathrm{CMB}}{T_i\,a_i}\right)^4,\label{eq:kapparad}\\
    \kappa_{\mathrm{s},\,i} & \equiv \kappa_{10}\cdot\left(\frac{10\,\mathrm{MeV}}{T_i}\right)^4 \left(\frac{a_{10}}{a_i}\right)^6, \label{eq:dd0_definition}
\end{align}
where $T_i=27\times10^9\,$K is the initial temperature
of \verb|AlterBBN| calculations \cite{2018arXiv180611095A},
$T_\mathrm{CMB}=2.7255$\,K is the observed temperature
of the cosmic microwave background (CMB) today \cite{2009ApJ...707..916F},
$a_i$ and $a_{10}$ denote the scale factors
at $T_i$ and $T_\mathrm{10}$, respectively.
These parameters measure the energy density ratios of each BSM component
to photons at $T_i$ when \verb|AlterBBN| starts its integration.
Eqs.~(\ref{eq:kapparad}) and (\ref{eq:dd0_definition})
are used to convert between $(\Delta N_\mathrm{eff},\,\kappa_{10})$
and $(\kappa_{\mathrm{rad},\,i},\,\kappa_{\mathrm{s},\,i})$ for \verb|AlterBBN|.
Our modified BBN codes are publicly available
at the links in Refs.~\cite{par_stiff,alter_stiff}.

In summary, we consider four free physical parameters,
$\Omega_\mathrm{b} h^2$, $\tau_n$, $\Delta N_\mathrm{eff}$ and $\kappa_\mathrm{10}$.
Their ranges are listed in Table~\ref{tab:priors}.
The bounds on $\Omega_\mathrm{b} h^2$ and $\tau_n$ are chosen
in accordance with their current measured values
\cite{2020A&A...641A...6P,2021PhRvL.127p2501G}.
The range of $\Delta N_\mathrm{eff}$ formally allows for
negative values for completeness.
Negative $\Delta N_\mathrm{eff}$ might occur in nonstandard scenarios, e.g.,
in which the temperature of the SM neutrinos is lower than expected.


\begin{table}[h]
\centering
\caption{Free parameters of BBNet and their ranges.
Eqs.~(\ref{eq:eta_definition}), (\ref{eq:kapparad}) and (\ref{eq:dd0_definition})
are used to convert these parameters to the intrinsic input of PArthENoPE and AlterBBN
when necessary.}
\label{tab:priors}
\begin{tabular}{l c}
\toprule
\textbf{Parameter} & \textbf{Range} \\
\midrule
$\Omega_\mathrm{b} h^2$ & $[0.005,\, 0.1]$ \\
$\tau_n$ (s) & $[875.5,\, 884.5]$ \\
$\Delta N_\mathrm{eff}$ & $[-1.0,\, 1.0]$ \\
$\kappa_\mathrm{10}$ & $[0,\, 1000]$ \\
\bottomrule
\end{tabular}
\end{table}

For illustration purposes, we present the dependences
of the predicted $Y_\mathrm{P}$ and $\mathrm{D/H}$
on $\kappa_{10}$ and $\Delta N_\mathrm{eff}$ in Fig.~\ref{fig:bbn_code_comparison}.
The results are based on our modified BBN codes, \verb|PArthENoPE| and \verb|AlterBBN|. 
The coincidence of the two curves in each panel
shows that the two solvers yield nearly identical primordial abundances
across the explored parameter ranges,
demonstrating the consistency between our implementations of the same BSM physics
in the two codes.

\begin{figure*}[t]
\centering
\includegraphics[width=1.0\textwidth]{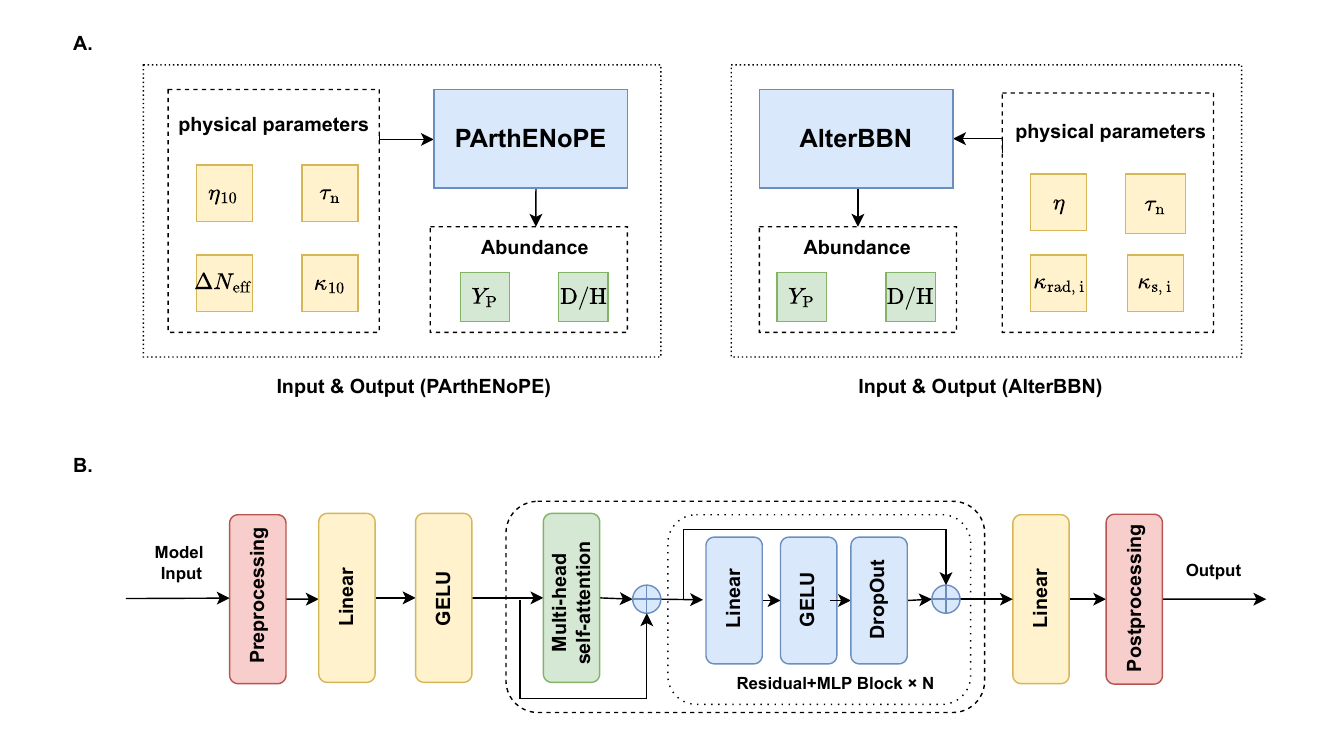}
\caption{Overview of the BBNet emulator and its training interface.
Panel A shows the input–output schematic for the two BBN solvers used to generate the training set. Each solver takes as input the baryon density $\Omega_{\mathrm{b}} h^{2}$, the neutron lifetime $\tau_{n}$, and any beyond-standard-model extensions. For scenarios with a stiff-fluid sector, the solvers parameterize this component through $\kappa_{10}$, defined as the stiff-fluid to photon energy-density ratio at $T = 10\,\mathrm{MeV}$. The solvers encode relativistic-sector modifications either through $\Delta N_{\mathrm{eff}}$ or through generalized coefficients $\kappa_{\mathrm{rad},i}$. They then return the primordial abundances $Y_{\mathrm{P}}$ and $\mathrm{D/H}$ for comparison with observations. Panel B depicts the BBNet architecture. The network projects the inputs with a linear–GELU layer, applies a multi-head self-attention module, and processes the result through $N$ residual MLP blocks. It finally uses a linear head to predict $(Y_{\mathrm{P}}, \mathrm{D/H})$.
}\label{fig:architecture}
\end{figure*}

\section{Methodology: Data Generation and Emulator Design}
\label{sec:methodology}

The \verb|BBNet| framework offers a flexible and standardized pipeline
for training BBN emulators that predict primordial element abundances,
regardless of the underlying physical model or the choice of BBN codes for generating the training data.
The pipeline follows a systematic workflow composed of data generation, training,
and inference (using the trained model to make predictions).

Operationally, the implementation mainly involves two stages.
As illustrated in Fig.~\ref{fig:architecture}A, the user first selects a desired physical model
and a preferred numerical BBN code to generate a training data set,
which essentially records the mapping between the physical parameters and the primordial element abundances;
the details are described in Sec.~\ref{sec:data_generation_strategy} below. 
Then, the user runs a standardized training procedure to obtain the emulator instance,
i.e., the trained neural network model with converged weight files particular to this set of training data.
This one-time data generation and training procedure is computationally inexpensive.
The neural network architecture of \verb|BBNet| is presented in Sec.~\ref{sec:nn_architecture}.
The link to the source code and a prescription for training emulator instances
is provided at the end of this paper.

During inference (Fig.~\ref{fig:architecture}B), \verb|BBNet| can output the primordial abundances
in milliseconds for a single set of input physical parameters. 
In a typical MCMC analysis, the user only needs to specify the pre-trained weight files
tied to the desired physical model and the BBN code,
without modifying the MCMC framework itself.

In this work, the default \verb|BBNet| emulators are based on the extended cosmological model
described in Sec.~\ref{sec:bbn_codes} and trained on the correspondingly modified BBN codes.
They yield predictions of the two best-measured primordial abundances, $Y_\mathrm{P}$ and $\mathrm{D/H}$. 

\subsection{Data generation strategy}
\label{sec:data_generation_strategy}

We generate separate training and validation data sets
using each of the two BBN codes, \verb|PArthENoPE| and \verb|AlterBBN|.
They are treated as the ground truth for our neural networks.
We sample the four physical parameters listed in Table~\ref{tab:priors},
assuming uniform priors on $\Omega_\mathrm{b}h^2$, $\Delta N_{\rm eff}$ and $\tau_n$,
and a log-uniform prior on $\kappa_{10}$ \footnote{In this work,
we adopt $\log_{10}\kappa_{10}\in[-7,\,3]$ with a uniform prior.}.
To ensure uniform sampling in the multidimensional parameter space,
we use the Latin hypercube sampling method \cite{Stein1987,Tang1993}.
The parameters are subsequently converted to the intrinsic input
of \verb|PArthENoPE| and \verb|AlterBBN| when necessary,
using Eqs.~(\ref{eq:eta_definition}), (\ref{eq:kapparad}) and (\ref{eq:dd0_definition}).
The data generation pipeline is illustrated in Fig.~\ref{fig:architecture}A.


All BBN computations are performed on Intel Xeon Gold CPUs.
For \verb|PArthENoPE|, we adopt its ``complete'' nuclear reaction network
(26 nuclides, 100 reactions) to generate the data set \cite{2008CoPhC.178..956P},
which costs $\sim 20-30\,$s (wall time, hardware–dependent) for a single run.
For \verb|AlterBBN|, we choose an ODE solver
based on the second-order Runge-Kutta algorithm with half step test
(setting ``failsafe=7'' in \verb|AlterBBN|; see \cite{2018arXiv180611095A}),
dubbed ``RK2\_halfstep'' throughout the paper.
This algorithm results in an average wall time per run of $\gtrsim 15$\,s.
The final data set contains 20,000 samples for each BBN code.
Note that the distributions of the two data sets are different from each other,
because a considerable part of the parameter space
(specified by Table~\ref{tab:priors}) is not accepted by \verb|PArthENoPE| as input,
while \verb|AlterBBN| accepts all combinations
of the parameters within their prior ranges.
For both data sets, we adopt the standard $8:1:1$ split
to obtain training, validation, and test data.


\subsection{Neural network architecture and data processing}
\label{sec:nn_architecture}

\verb|BBNet| is designed to ingest the four-dimensional vector
of $(\Omega_\mathrm{b}h^2,\,\tau_n,\,\Delta N_{\rm eff},\,\kappa_{10})$
and output the primordial element abundances, $(Y_\mathrm{P},\,\mathrm{D/H})$.
Its overall neural network architecture is illustrated in Fig.~\ref{fig:architecture}B.
All input and output variables are transformed and standardized
to have zero mean and unit variance before training,
indicated by the preprocessing and postprocessing modules.
These data projection techniques help \verb|BBNet|
learn patterns in the dependence of output on input with balanced weights,
hence achieving a more efficient training process
and increasing the accuracy of the resultant model.




As shown in Fig.~\ref{fig:architecture}B,
the preprocessed input data are reshaped into 4096-dimensional vectors
via a linear projection.
They are then passed through a Gaussian Error Linear Unit (GeLU)
activation function \cite{Hendrycks2016}.
These two layers can be expressed together as
\begin{equation}
    h_{0} = \mathrm{GeLU}(W_{\mathrm{lin}} X_{\mathrm{scaled}} + b_{\mathrm{lin}}),
\end{equation}
where $W_{\mathrm{lin}}$ and $b_{\mathrm{lin}}$
are learnable weights and bias parameters in the linear projection layer.

We employ a multi-head self-attention (MHSA) module
to further reparameterize the data \cite{Vaswani2017}.
The MHSA module can dynamically rescale
the relative contributions of the physical parameters. 
We use eight attention heads, with the MHSA operation defined as
\begin{equation}
    \mathrm{MHSA}(h_{0}) = \mathrm{Concat}(\mathrm{head}_1, \dots, \mathrm{head}_8)\,W_{O},
\end{equation}
where each head is computed as
\begin{equation}
    \mathrm{head}_i = \mathrm{softmax}\!\left(\frac{Q_i K_i^\top}{\sqrt{d_k}}\right)V_i.
\end{equation}
Here $Q_i$, $K_i$, and $V_i$ denote the linear projections learned from the input $h_{0}$, and $d_k$ is the key dimension, 
included as a scaling factor to stabilize the softmax layer \cite{Vaswani2017}.
The MHSA module is implemented with a residual connection:
\begin{align}
    h_{1} = h_{0} + \mathrm{MHSA}\!\left(h_{0}\right).
\end{align}
We then take $h_{1}$ to initialize
the residual multilayer perceptron (MLP) stack with $N$ blocks, expressed as:
\begin{equation}
    h_{t+1} \;=\; h_{t} \;+\;
    \mathrm{Dropout}\,\Big(\mathrm{GELU}\big(W_t h_t + \mathbf{b}_t\big)\Big),
\end{equation}
where $W_t$ and $\mathbf{b}_t$ are the learnable parameters of the $t$-th MLP block,
$t=1,\dots,N$.
A dropout rate of $0.3$ is used to regularize training and mitigate overfitting
\cite{Srivastava2014}.

Finally, another linear projection layer transforms the 4096-dimensional data vectors
to the predicted abundances $(Y_\mathrm{P},\,\mathrm{D/H})$,
up to a postprocessing step.

\section{BBNet Training and Validation}
\label{sec:training}

This section details the training strategy of \verb|BBNet|,
which is proven to be successful and stable
on both data sets (generated by \verb|PathENoPE| and \verb|AlterBBN|).
Our techniques here can be generalized to train emulators for other BBN codes.

\subsection{Optimizer and scheduler}
\label{sec:optimization}

The \verb|BBNet| model is trained using the AdamW optimizer.
It combines the adaptive learning rate advantages
of the Adam optimizer with decoupled weight decay
\cite{KingmaBa2015Adam,Loshchilov2019}.
For both data sets, the initial learning rate is set to $5 \times 10^{-5}$,
with a weight decay parameter $\lambda_W = 10^{-5}$.
To avoid exploding gradients,
we apply gradient clipping with a maximum norm threshold of 1.0.

We adopt a batch size of 16 for each iteration of the training processes.
This batch size balances computational costs (GPU memory) and model complexity.
We employ a scheduler that reduces the learning rate by a factor of 0.5
whenever the validation loss stops improving for 10 consecutive epochs.
This strategy promotes rapid initial convergence
and allows for finer parameter adjustments once the model begins to plateau.

\subsection{Regularization}
\label{sec:regularization}

During training and validation, we apply several regularization techniques
to mitigate overfitting and improve the generalizability of the model to unseen data.
Dropout is implemented by randomly setting a fraction of activations to zero.
It discourages complex coadaptations between neurons,
so that the model is inclined to learn more robust features.
Weight decay is integrated through the AdamW optimizer, as mentioned above.
This operation further enhances the stability of the model
by suppressing large weights and favoring simpler representations.

We use the standard mean absolute error (MAE) for the loss function.
We also include a regularization term
that enforces the local stability of model predictions
under small perturbations of the input parameters.
It penalizes the differences between the model outputs
at $\mathbf{x}$ and $\mathbf{x}{+}\boldsymbol{\epsilon}$
(here $\boldsymbol{\epsilon}$ denotes a Gaussian perturbation),
expressed as follows:
\begin{equation}
  \mathcal{L}_{\mathrm{smooth}}
  = \mathbb{E}_{\boldsymbol{\epsilon}\sim\mathcal{N}(\mathbf{0},\,\sigma^{2}\mathbf{I})}
  \,\left\|
    f(\mathbf{x}{+}\boldsymbol{\epsilon})
    - f(\mathbf{x})
  \right\|_{2}^{2},
\end{equation}
where $\|\cdot\|_{2}$ is the Euclidean norm of the output vector
and $\sigma=0.02$ is fixed in all reported experiments.
This term suppresses fluctuations in the learned mapping,
guaranteeing smooth responses to input parameters.

In addition, we introduce another degree of regularization
in terms of the following symmetric mean absolute percentage error (sMAPE)
concerning the deuterium abundance:
\begin{equation}
  \mathcal{L}_{\mathrm{sMAPE}}
  = \frac{1}{N}\sum_{i=1}^{N}
  \frac{2\,\big|\hat{y}^{(\mathrm{raw})}_{\mathrm{D/H},i}
                - y^{(\mathrm{raw})}_{\mathrm{D/H},i}\big|}
       {\big|\hat{y}^{(\mathrm{raw})}_{\mathrm{D/H},i}\big|
        + \big|y^{(\mathrm{raw})}_{\mathrm{D/H},i}\big|}.
\end{equation}
This term helps the model achieve the desired accuracy
in the predicted values of D/H, 
which span many orders of magnitude in our data sets.

\begin{figure}[t]
  \centering
  \subfloat[PArthENoPE\label{fig:train_val_loss_parthenope}]{%
    \includegraphics[width=0.48\columnwidth]{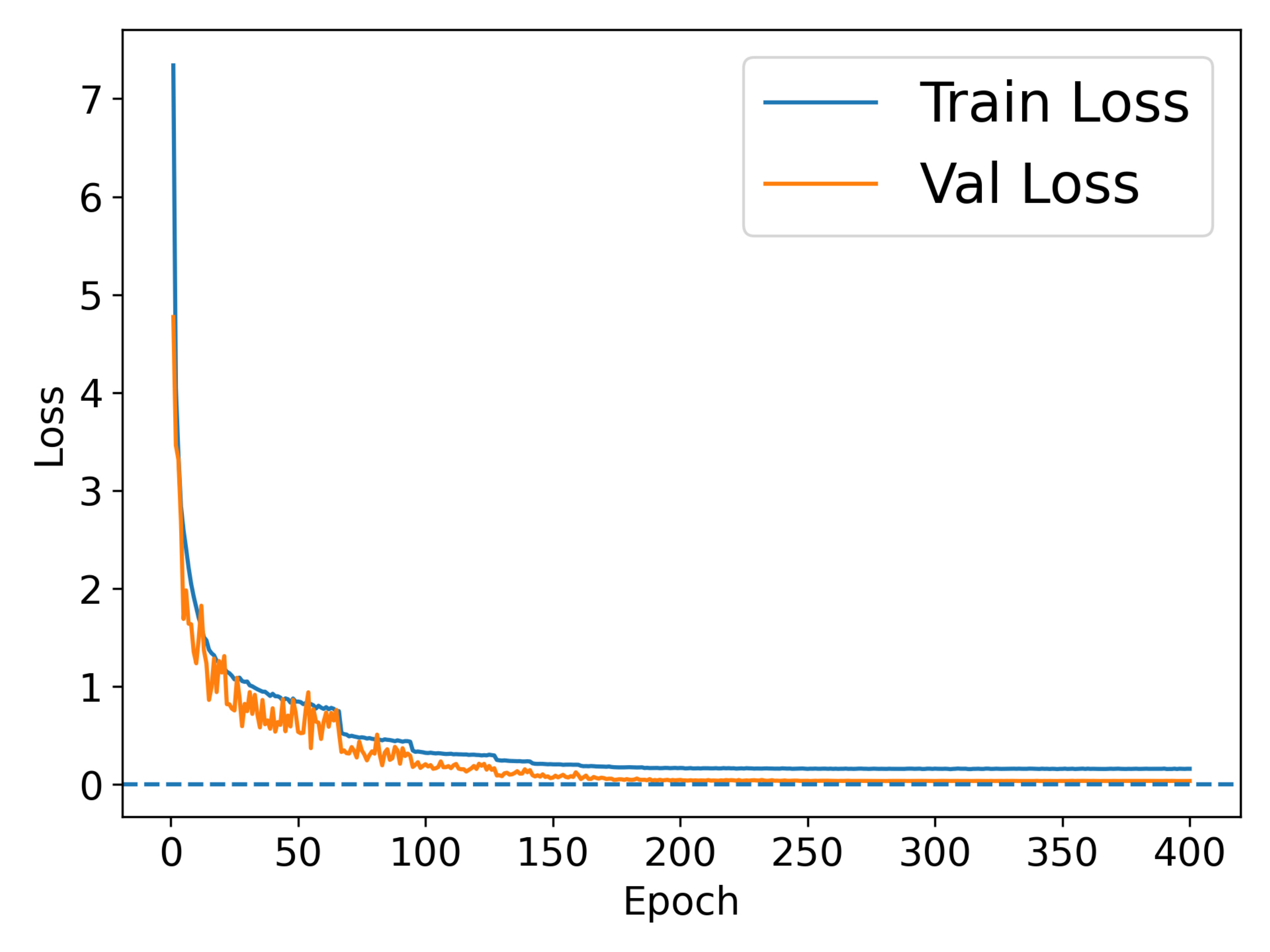}
  }
  \hfill
  \subfloat[AlterBBN\label{fig:train_val_loss_alterbbn}]{%
    \includegraphics[width=0.48\columnwidth]{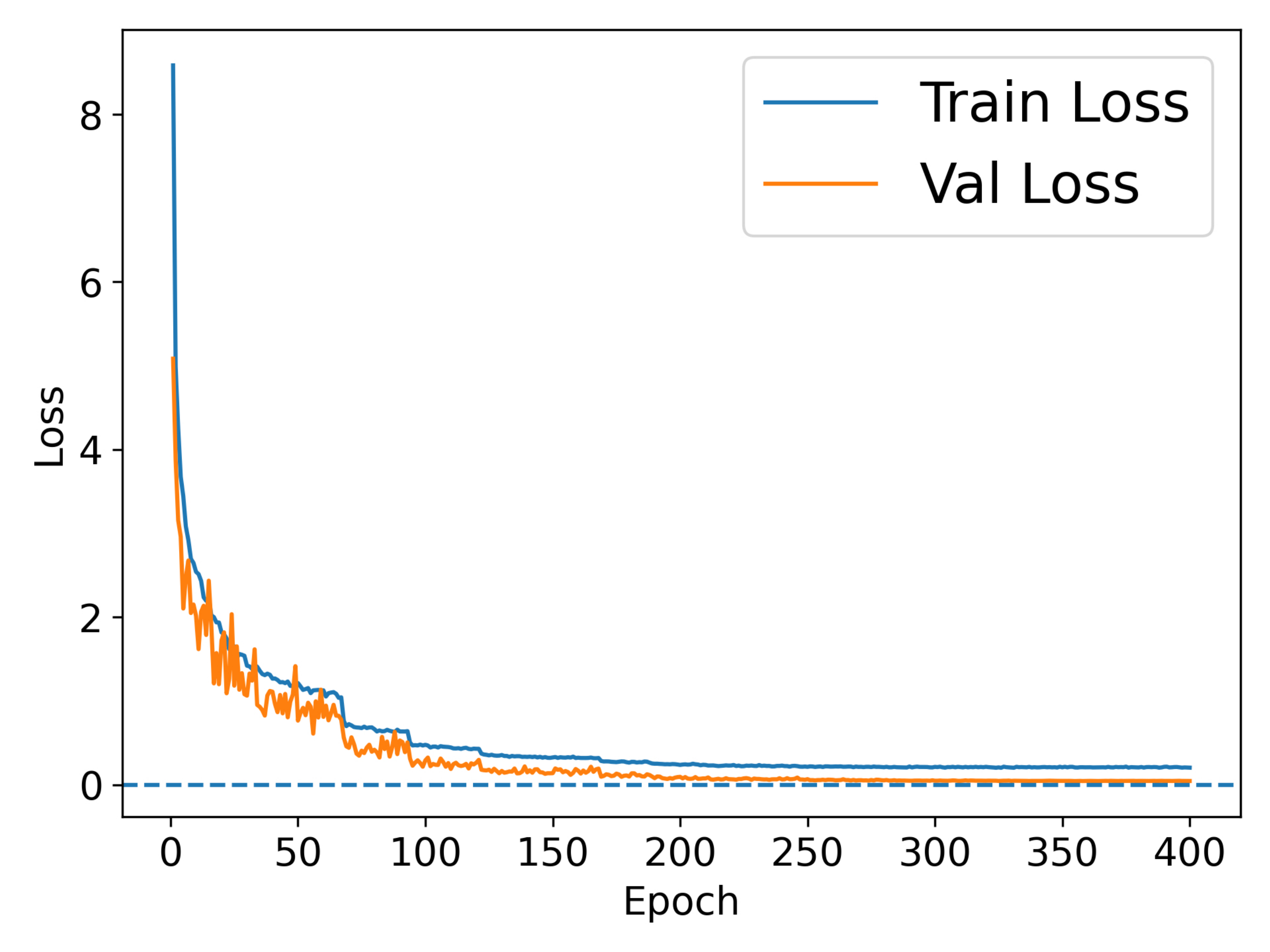}
  }
  \caption{Training and validation loss curves for the BBNet emulators.
  Figs.~\ref{fig:train_val_loss_parthenope} and \ref{fig:train_val_loss_alterbbn}
  illustrate the training processes of the neural network models
  based on data sets obtained by our modified PArthENoPE and AlterBBN, respectively.
  The loss function is defined as the mean absolute error
  between the predicted and true values of $(Y_{\mathrm{P}},\,\mathrm{D/H})$.
  Both losses decrease rapidly and stabilize within 300 epochs, 
  reaching values close to zero with no indications of overfitting.}
  \label{fig:train_val_loss_combined}
\end{figure}

\subsection{Training and validation loss}
\label{sec:validation}

The total objective function combines the MAE loss
(weighted between $Y_\mathrm{P}$ and $\mathrm{D/H}$)
and the two regularization terms described above.
It is formally written as
\begin{equation}
  \mathcal{L}_{\mathrm{total}}
  = \mathcal{L}_{\mathrm{MAE}}
  + \lambda_{\mathrm{smooth}}\,\mathcal{L}_{\mathrm{smooth}}
  + \lambda_{\mathrm{D/H}}\,\mathcal{L}_{\mathrm{sMAPE}},
\end{equation}
where $\lambda_{\mathrm{smooth}}$ and $\lambda_{\mathrm{D/H}}$ are hyperparameters.
We set $\lambda_{\mathrm{smooth}}=0.1$ and $\lambda_\mathrm{D/H}=50$.

The loss curves for training and validation are presented
in Fig.~\ref{fig:train_val_loss_combined} for both \verb|BBNet| models.
The convergence of the two training processes
confirms the stability of the neural network models
and the generalizability of our framework.
The closeness of the training and validation curves for both data sets
demonstrates that our \verb|BBNet| models have achieved optimal complexity.

\section{Results}
\label{sec:results}


We perform a comprehensive twofold evaluation of the trained \verb|BBNet| emulators
to verify both its predictive accuracy and computational efficiency.
In Sec.~\ref{sec:accuracy_comparison}, we quantitatively benchmark
the emulator predictions against the ground-truth outputs of the underlying BBN codes.
We also compare them with the current observational constraints
on $Y_\mathrm{P}$ and $\mathrm{D/H}$.
Subsequently, Sec.~\ref{app:parameter_dependence} explores
the dependence of the predicted primordial abundances
on the physical parameters over their most relevant ranges.


\subsection{Emulation accuracy and efficiency}
\label{sec:accuracy_comparison}

\begin{figure}[h]
  \centering
  \subfloat[PArthENoPE\label{fig:accuracy_pe}]{%
    \includegraphics[width=0.9\columnwidth]{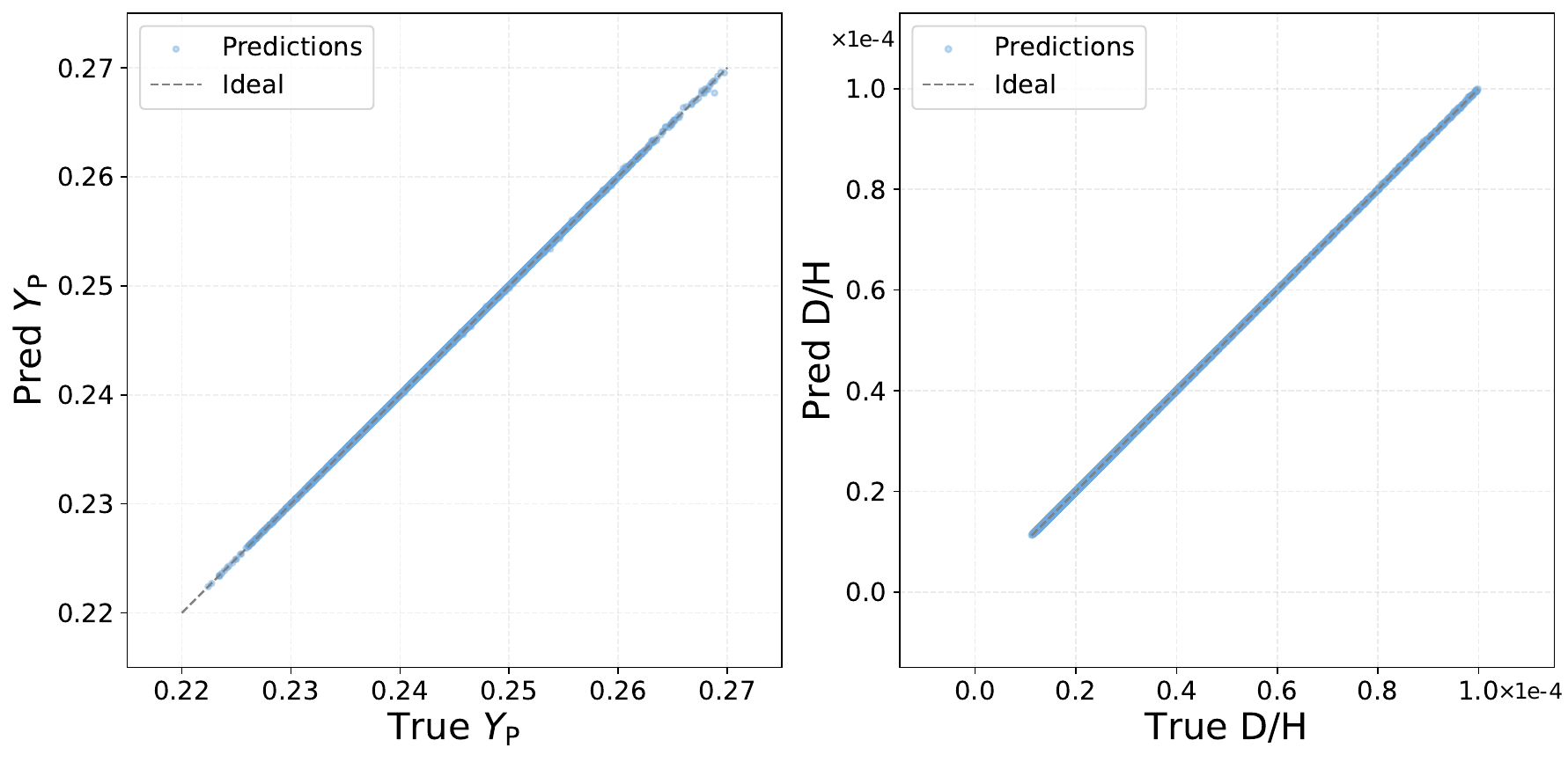}
  }
  \vspace{0.5em}
  \subfloat[AlterBBN\label{fig:accuracy_alterbbn}]{%
    \includegraphics[width=0.9\columnwidth]{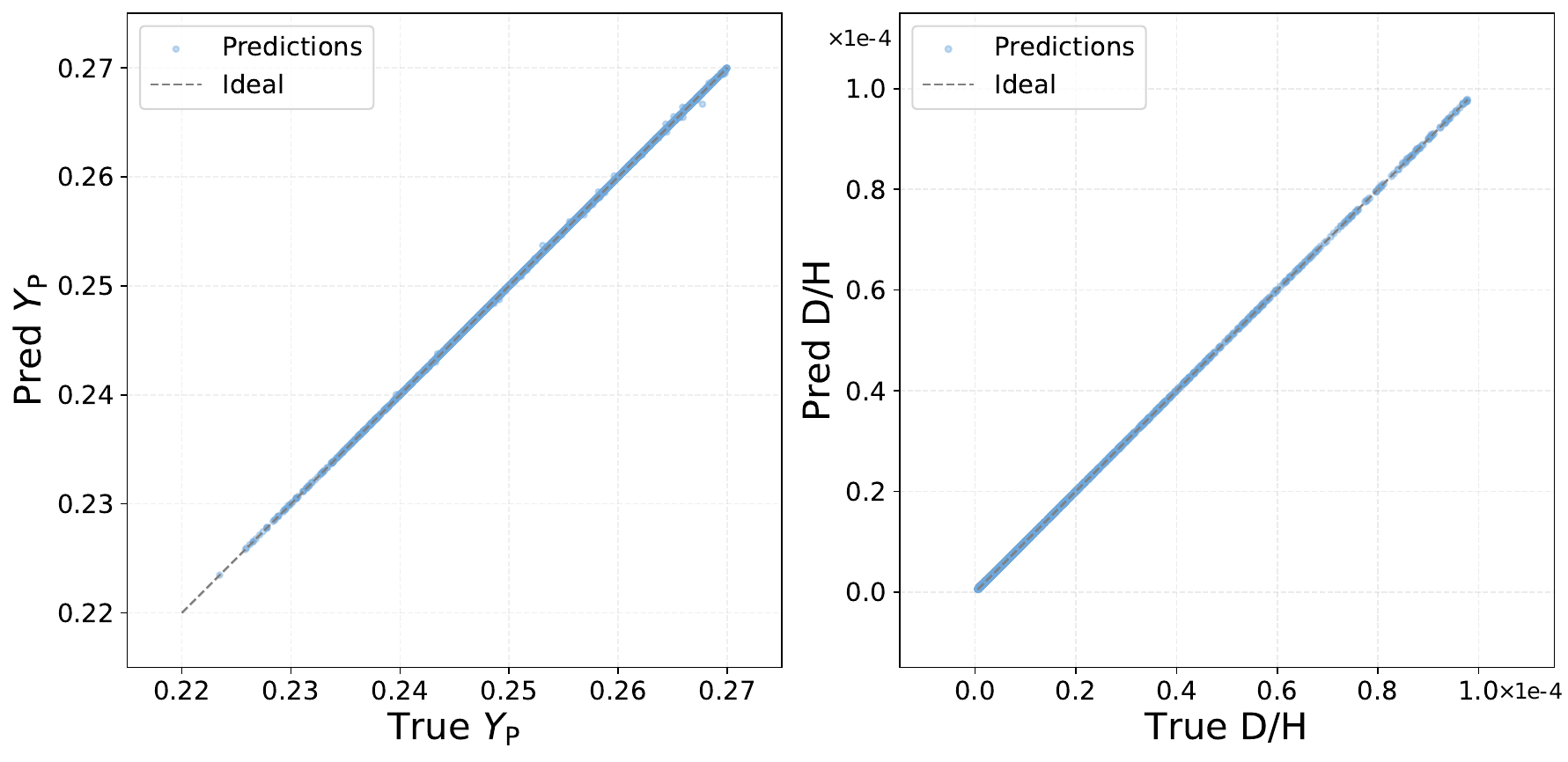}
  }
  \caption{Comparison between the primordial element abundances
  predicted by BBNet and the ground truth.
  Each panel compares BBNet predictions on the vertical axis
  with outputs of the benchmark solver on the horizontal axis.
  Fig.~\ref{fig:accuracy_pe} shows results based on PArthENoPE
  using its complete network. Fig.~\ref{fig:accuracy_alterbbn}
  shows results based on AlterBBN with its RK2\_halfstep mode.
  Each subpanel presents $Y_\mathrm{P}$ on the left and $\mathrm{D/H}$ on the right, where blue scatter points denote individual test samples plotted against the corresponding solver values.
  The diagonal dashed line in each plot indicates ideal emulation with $\hat{y} = y$.}
  \label{fig:bbnet_accuracy_combined}
\end{figure}

We first assess the performance of \verb|BBNet|
by its accuracy in reproducing the results of BBN calculations.
For both test sets generated by \verb|PArthENoPE| and \verb|AlterBBN|,
the \verb|BBNet| emulator accurately reproduces
the true values of $Y_\mathrm{P}$ and $\mathrm{D/H}$,
as illustrated in Fig.~\ref{fig:bbnet_accuracy_combined}.
Note that \verb|AlterBBN| can output small $\mathrm{D/H}$ values $(<10^{-5})$
while \verb|PArthENoPE| only allows for
values of the same order as the measured deuterium abundance
(see the upper right panel of Fig.~\ref{fig:bbnet_accuracy_combined}).
Hence, to achieve the desired accuracy
in predicting small values of $\mathrm{D/H}$ produced by \verb|AlterBBN|,
our corresponding \verb|BBNet| emulator
trains distinct models (expert modes) for two ranges of $\mathrm{D/H}$,
$[10^{-7},\,10^{-5}]$ and $[10^{-5},\,10^{-3}]$, respectively
(see Appendix~\ref{experts} for details).

In all cases illustrated in Fig.~\ref{fig:bbnet_accuracy_combined},
the close agreement between predictions and ground-truth values
demonstrates that the \verb|BBNet| framework is not confined
to the internal implementation of any single BBN solver.
Instead, it effectively learns the underlying functional dependence
between the physical parameters and primordial light element abundances.

To further quantify this agreement, we evaluate two complementary regression metrics
that characterize both the magnitude and the direction of residual errors.
First, the mean percentage error (MPE) measures the average signed relative difference
between the predicted and the true abundances, defined as
\begin{equation}
    \mathrm{MPE}
    = \frac{1}{N}\sum_{i=1}^{N}\frac{\hat{y}_i - y_i}{y_i} \times 100\%.
\end{equation}
We also consider the root mean square percentage error (RMSPE)
that describes the absolute magnitude of relative errors as follows:
\begin{equation}
    \mathrm{RMSPE}
    = \sqrt{\frac{1}{N}\sum_{i=1}^{N}
    \left(\frac{\hat{y}_i - y_i}{y_i}\times100\% \right)^{2}}.
\end{equation}

Regarding observational constraints on the primordial abundances,
we adopt the values reported by PDG 2024
as the benchmark here \cite{Fields:2025},
\begin{IEEEeqnarray}{rCl}\label{eq:obs_constraints}
    Y_\mathrm{P} & = & 0.245 \pm 0.003,\nonumber\\
    \mathrm{D/H} & = & (2.547 \pm 0.029)\times 10^{-5}.
\end{IEEEeqnarray}
These values were compiled from a group of astronomical measurements,
e.g., Refs.~\cite{2021JCAP...03..027A,2019ApJ...876...98V,2019MNRAS.487.3221F,2020ApJ...896...77H,2021MNRAS.505.3624V,2022MNRAS.510..373A,2021MNRAS.502.3045K} for $Y_\mathrm{P}$
and Refs.~\cite{2014ApJ...781...31C,2016ApJ...830..148C,2015MNRAS.447.2925R,2016MNRAS.458.2188B,2017MNRAS.468.3239R,2018JPhCS1038a2012Z,2018ApJ...855..102C} for $\mathrm{D/H}$.

We summarize the quantitative results of emulator performance
in Table~\ref{tab:bbnet_comparison} using the metrics defined above.
Both the RMSPE and the MPE values are far below
the observational uncertainties—by more than four orders of magnitude
for $Y_\mathrm{P}$—demonstrating that \verb|BBNet| accurately reproduces
the theoretical values of $Y_\mathrm{P}$ and $\mathrm{D/H}$ with negligible errors.
\verb|BBNet| is therefore a fully competent alternative to BBN codes
for parameter inference tasks.

\begin{table}[ht]
\caption{Comparison of RMSPE, MAPE, and MPE (\%) for \textsc{BBNet}
trained on two BBN solvers, evaluated against their respective ground truths.}
\label{tab:bbnet_comparison}
\begin{ruledtabular}
\begin{tabular}{lccc}
\textbf{Abundance} & \textbf{RMSPE (\%)} & \textbf{MAPE (\%)} & \textbf{MPE (\%)} \\
\hline
\multicolumn{4}{l}{\textit{PArthENoPE}}\\
$Y_\mathrm{P}$      & 0.0158 & 0.0064 & $-0.0011$ \\
$\mathrm{D/H}$      & 0.0503 & 0.0331 & $0.0013$ \\
\multicolumn{4}{l}{\textit{AlterBBN}}\\
$Y_\mathrm{P}$      & 0.0175 & 0.0055 & $-0.0014$ \\
$\mathrm{D/H}$      & 0.0799 & 0.0455 & $0.0002$ \\
\end{tabular}
\end{ruledtabular}
\end{table}

\begin{figure}[t]
  \centering
  \subfloat[PArthENoPE\label{fig:time_parthenope}]{
    \includegraphics[width=0.47\textwidth,keepaspectratio]{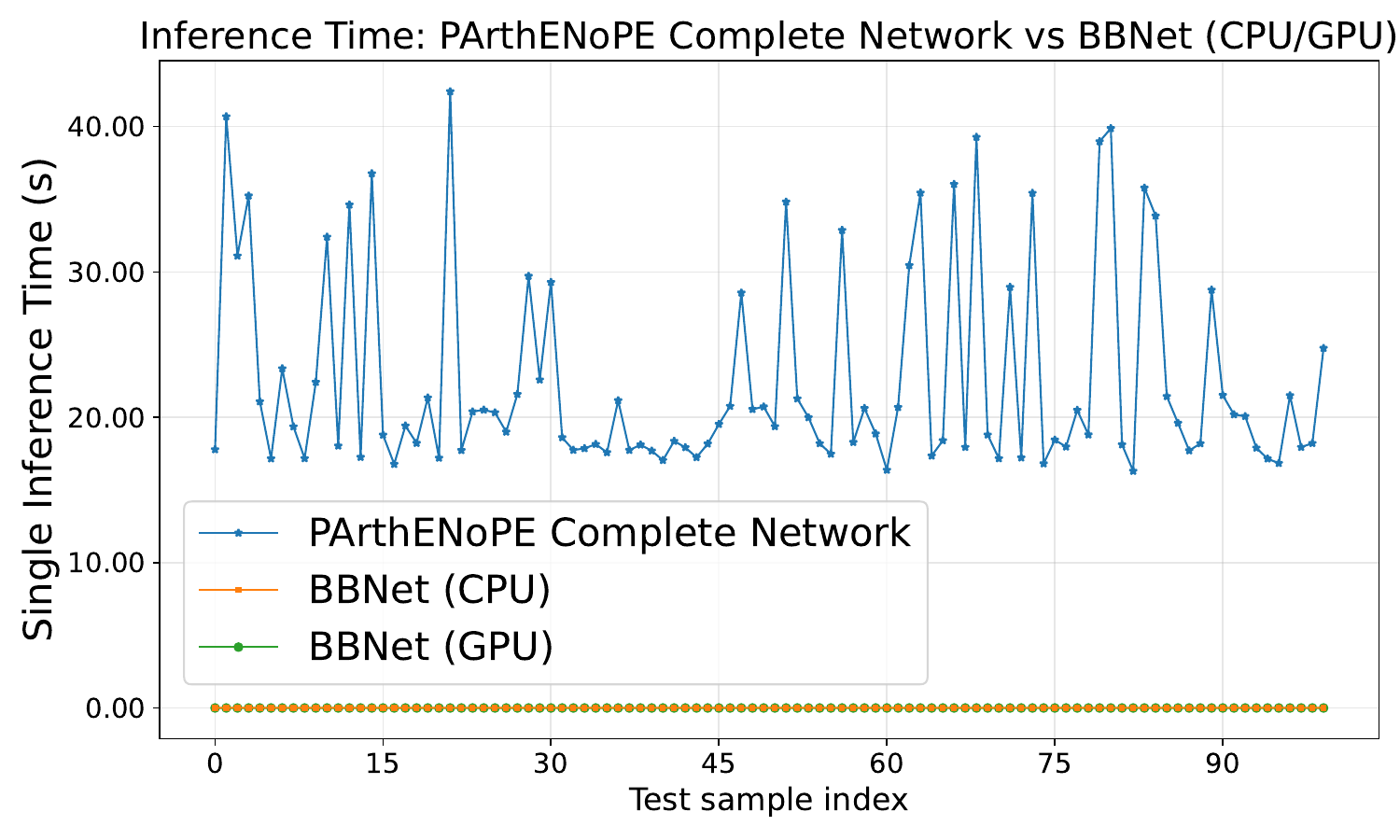}
  }
  \hfill
  \subfloat[AlterBBN\label{fig:time_alterbbn}]{
    \includegraphics[width=0.47\textwidth,keepaspectratio]{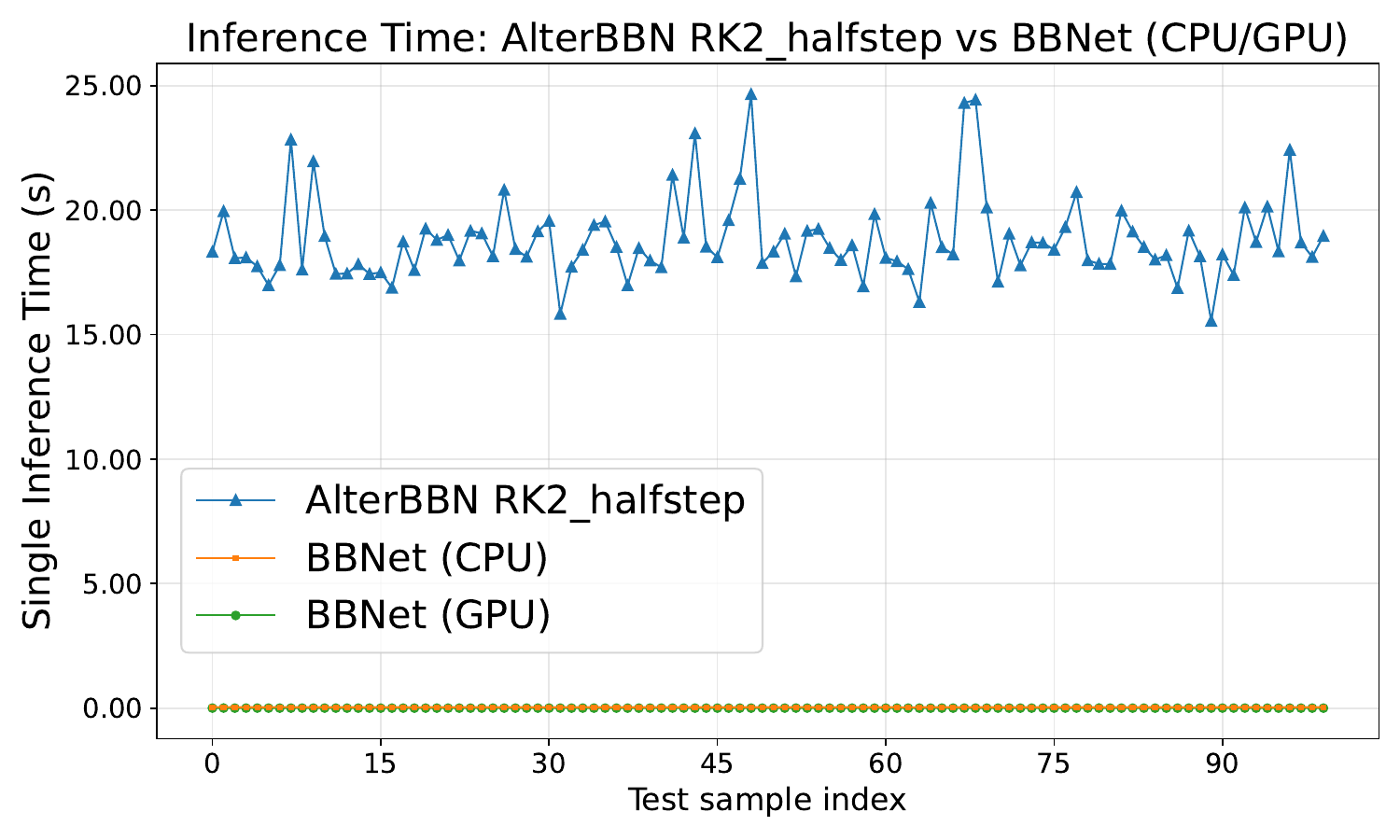}
  }
  \caption{Single-sample inference times for BBN numerical solvers
  and the BBNet emulator.
  The wall times are measured for 100 test samples for each method, with the horizontal axis denoting the sample index.
  Blue lines indicate the runtimes of the baseline solvers,
  namely the complete network PArthENoPE (Fig.~\ref{fig:time_parthenope})
  and AlterBBN with the RK2\_halfstep mode (Fig.~\ref{fig:time_alterbbn}).
  Orange and green lines show the corresponding inference times of the BBNet emulator executed on CPU and GPU, respectively.
  The linear scale on the $y$-axis helps visualize the huge difference between the inference times of traditional BBN codes and our neural network emulators.
  BBNet achieves a speed-up of $\sim 10^{3}-10^{4}$ times in both cases
  and attains sub-millisecond evaluations on GPU with minimal overhead on CPU.
  }
  \label{fig:inference_time_comparison}
\end{figure}

In addition to the accuracy of model predictions, computational efficiency
is equally critical for applications that require large-scale evaluations.
Fig.~\ref{fig:inference_time_comparison} compares
the latency (wall time) per execution between different methods.
On both CPUs and GPUs, \verb|BBNet| achieves an average speed-up
of several orders of magnitude relative to traditional ODE-based solvers.
This dramatic acceleration enables real-time parameter estimation using MCMC sampling,
where millions of abundance evaluations are typically required.

\begin{figure*}[ht]
  \centering
  \newlength{\panelw}\setlength{\panelw}{0.24\textwidth}
  \newlength{\panelh}\setlength{\panelh}{6cm}
  \subfloat[\label{fig:dependence_kappa_pe}]{%
    \begin{minipage}[t][\panelh][c]{\panelw}
      \centering
      \includegraphics[height=\panelh,keepaspectratio]{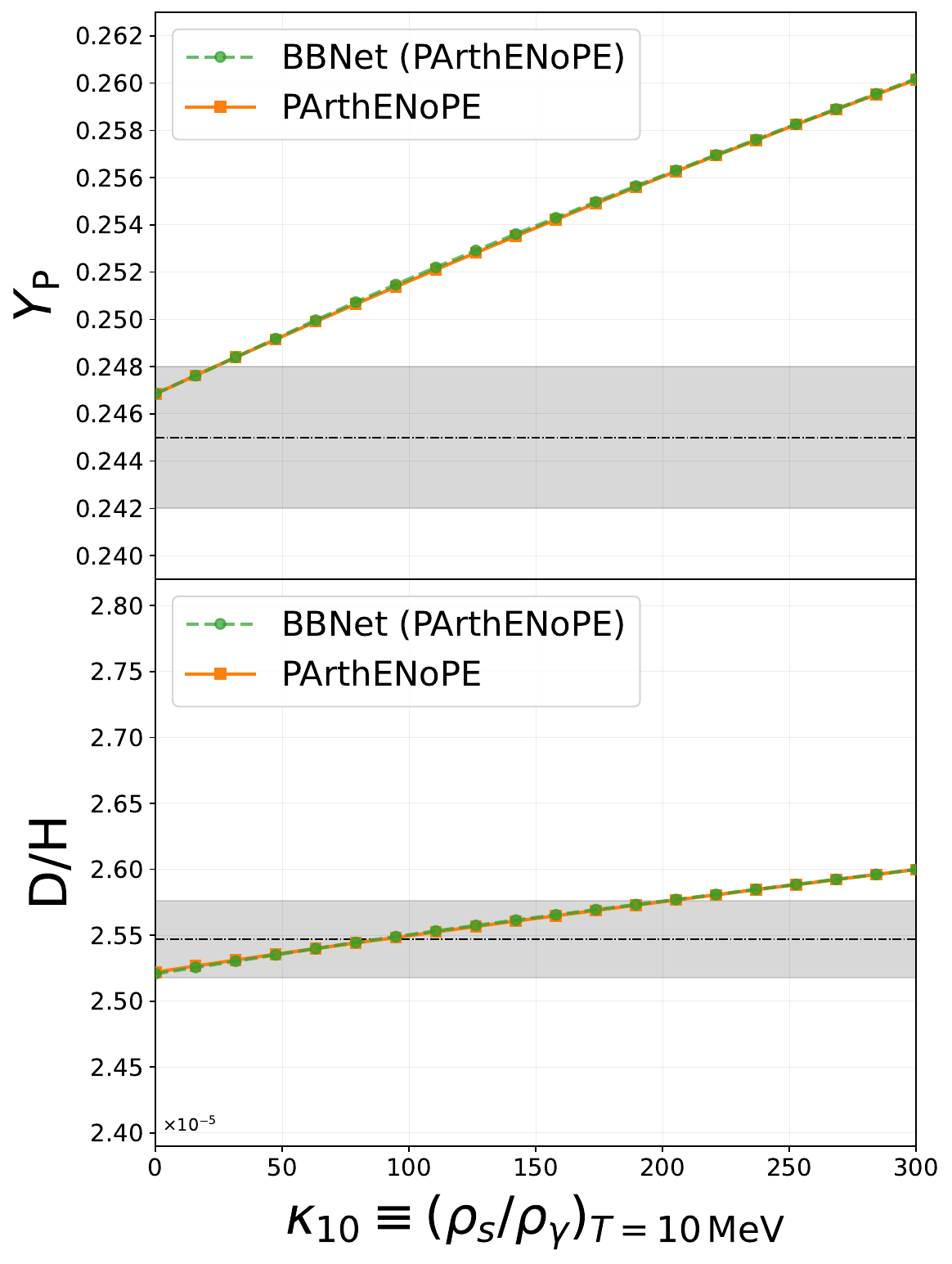}
    \end{minipage}
  }\hfill
  \subfloat[\label{fig:dependence_dnnu_pe}]{%
    \begin{minipage}[t][\panelh][c]{\panelw}
      \centering
      \includegraphics[height=\panelh,keepaspectratio]{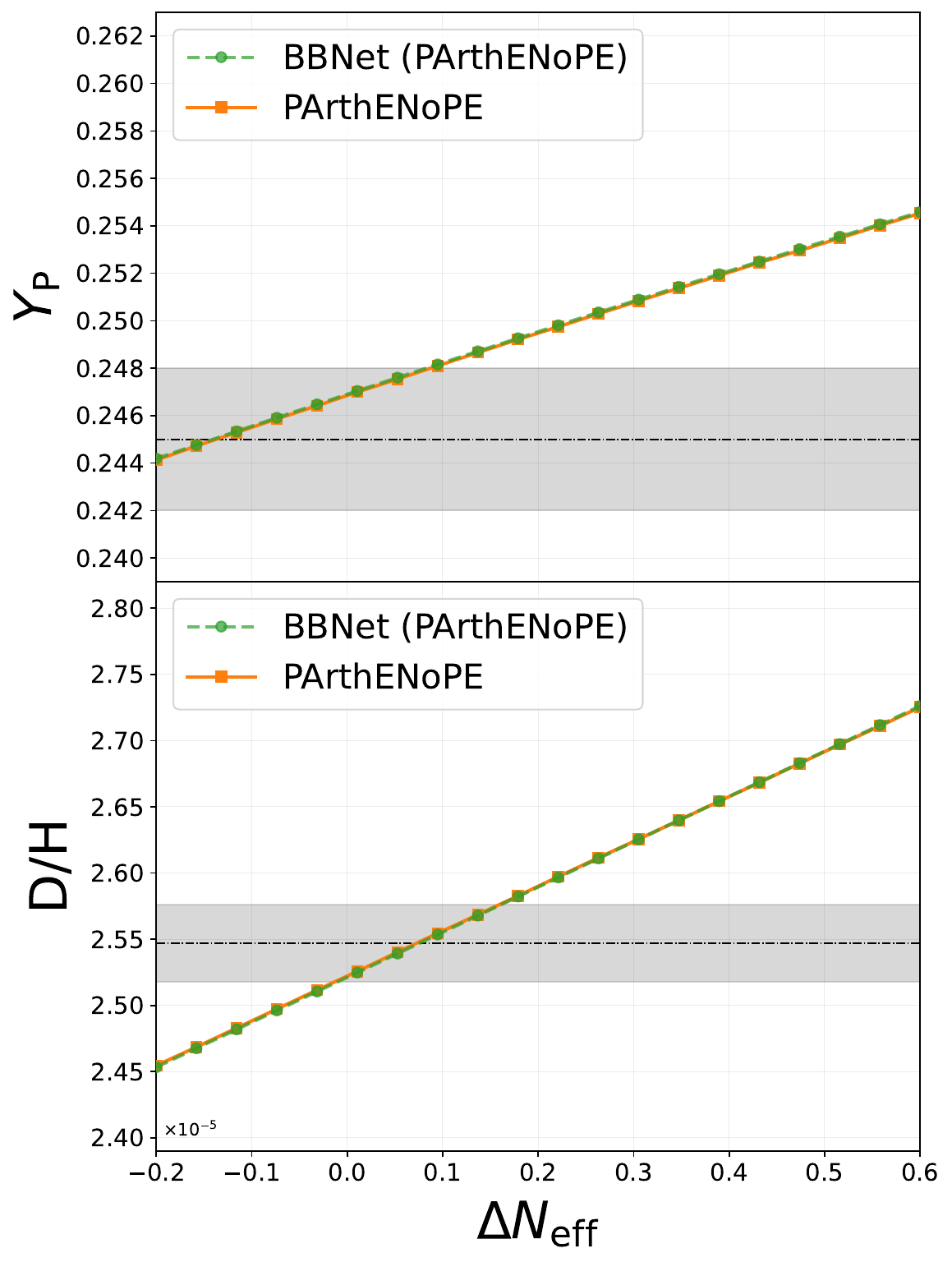}
    \end{minipage}
  }\hfill
  \subfloat[\label{fig:dependence_omega_pe}]{%
    \begin{minipage}[t][\panelh][c]{\panelw}
      \centering
      \includegraphics[height=\panelh,keepaspectratio]{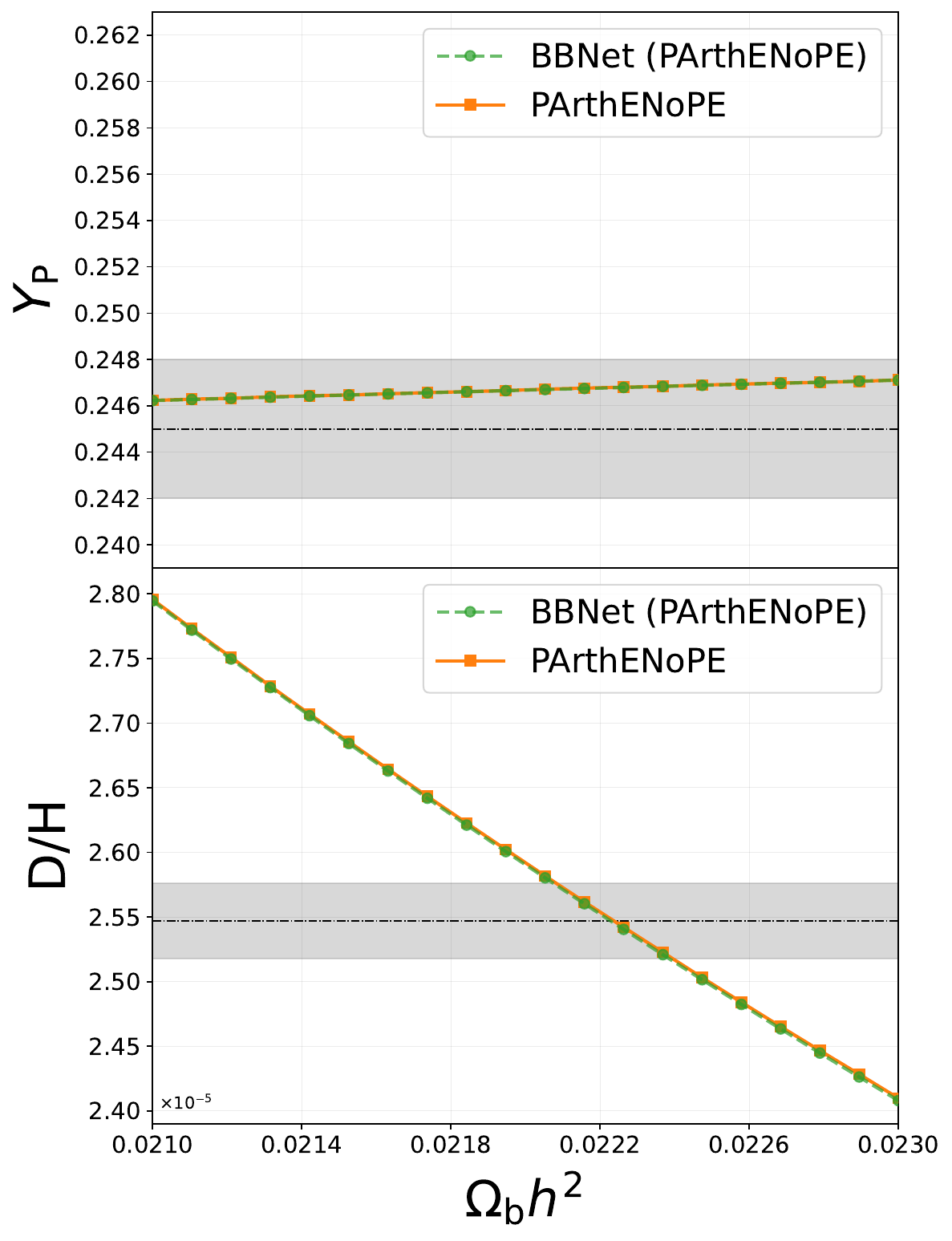}
    \end{minipage}
  }\hfill
  \subfloat[\label{fig:dependence_tau_pe}]{%
    \begin{minipage}[t][\panelh][c]{\panelw}
      \centering
      \includegraphics[height=\panelh,keepaspectratio]{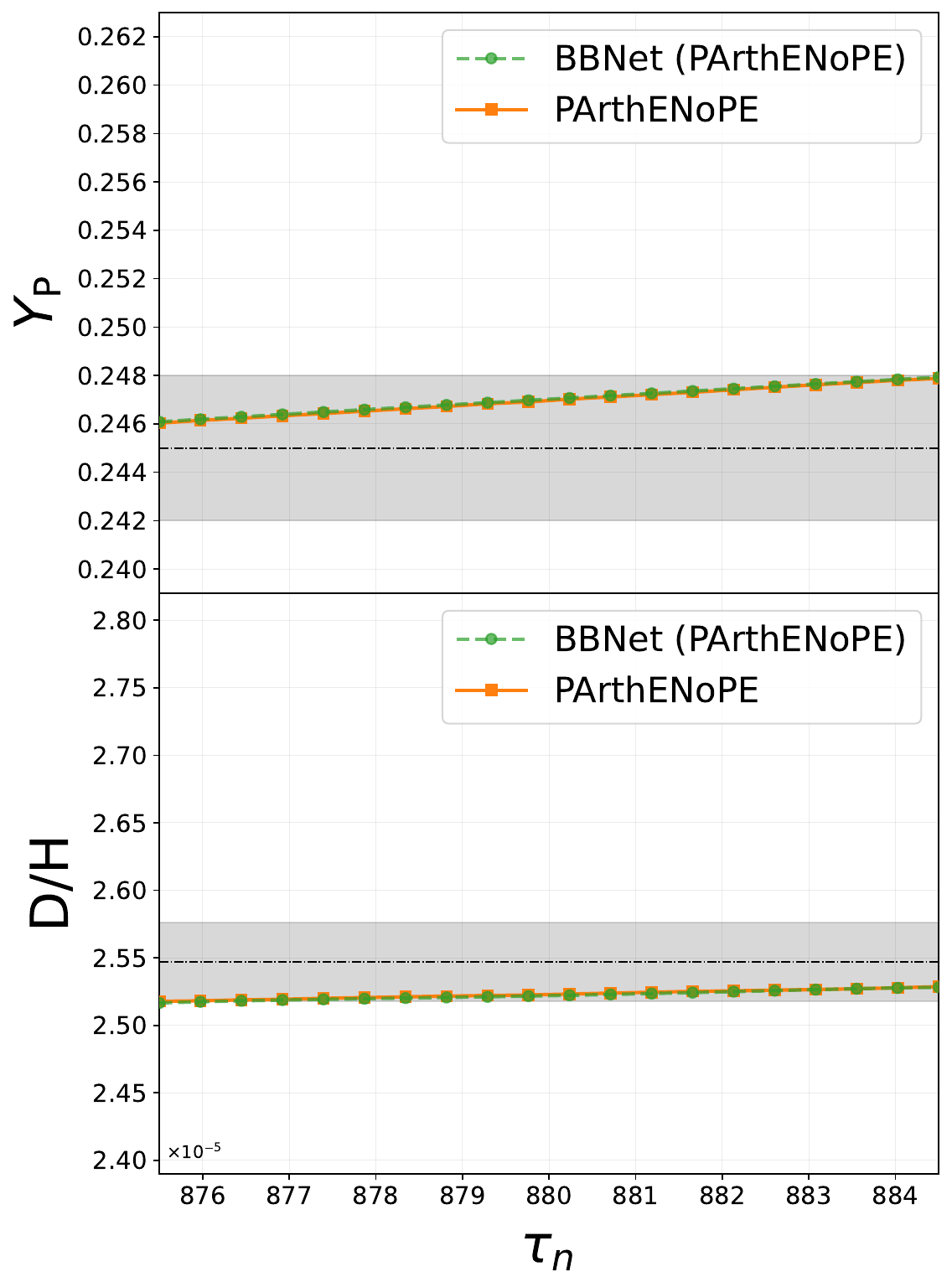}
    \end{minipage}
  }
  \caption{Parameter dependence of primordial abundances from PArthENoPE and BBNet.
  The top axes show the dependence of the primordial helium mass fraction $Y_{\mathrm{P}}$, and the bottom axes show the dependence of the deuterium-to-hydrogen ratio $\mathrm{D/H}$.
  Predictions from PArthENoPE appear as solid curves with circle markers,
  and the BBNet emulator appears as dashed curves with circle markers.
  Fig.~\ref{fig:dependence_kappa_pe} varies
  the density parameter of the stiff fluid, $\kappa_{10}$.
  Fig.~\ref{fig:dependence_dnnu_pe} varies $\Delta N_{\mathrm{eff}}$,
  Fig.~\ref{fig:dependence_omega_pe} varies
  the baryon density $\Omega_{\mathrm{b}} h^{2}$,
  and Fig.~\ref{fig:dependence_tau_pe} varies the neutron lifetime $\tau_{n}$.
  In each case, all remaining cosmological and nuclear inputs are held fixed at their fiducial values. Gray shaded bands show the one-sigma observational intervals adopted from PDG 2024 \cite{Fields:2025}.
  Black dashed lines indicate the corresponding central values.}
    \label{fig:appendix_dependence_parthenope}
\end{figure*}

\begin{figure*}[ht]
  \centering
  \newlength{\panelwB}\setlength{\panelwB}{0.24\textwidth}
  \newlength{\panelhB}\setlength{\panelhB}{6cm}

  \subfloat[\label{fig:sens:kappa_pe}]{%
    \begin{minipage}[t][\panelhB][c]{\panelwB}
      \centering
      \includegraphics[height=\panelhB,keepaspectratio]{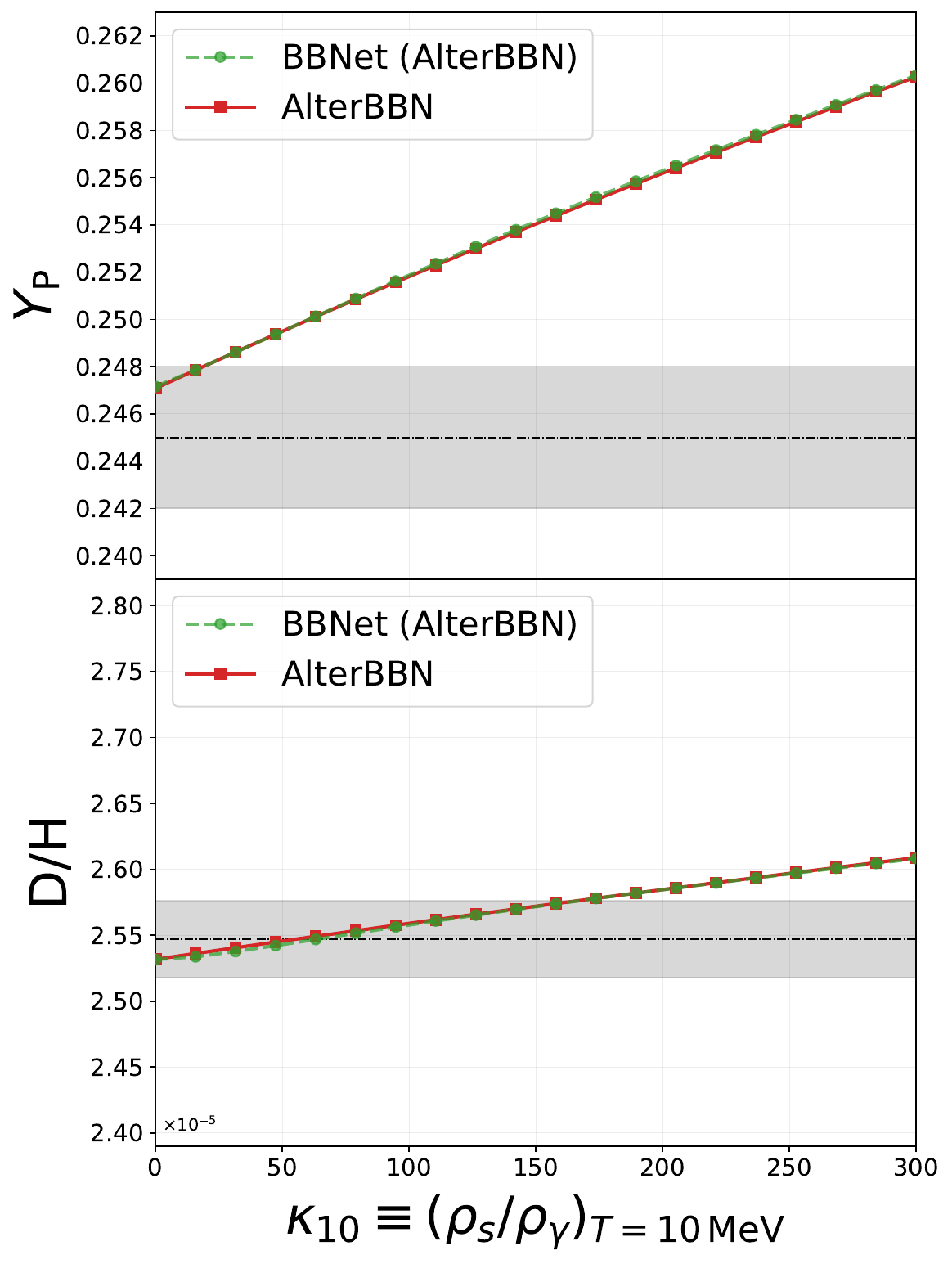}
    \end{minipage}
  }\hfill
  \subfloat[\label{fig:sens:dNnu_pe}]{%
    \begin{minipage}[t][\panelhB][c]{\panelwB}
      \centering
      \includegraphics[height=\panelhB,keepaspectratio]{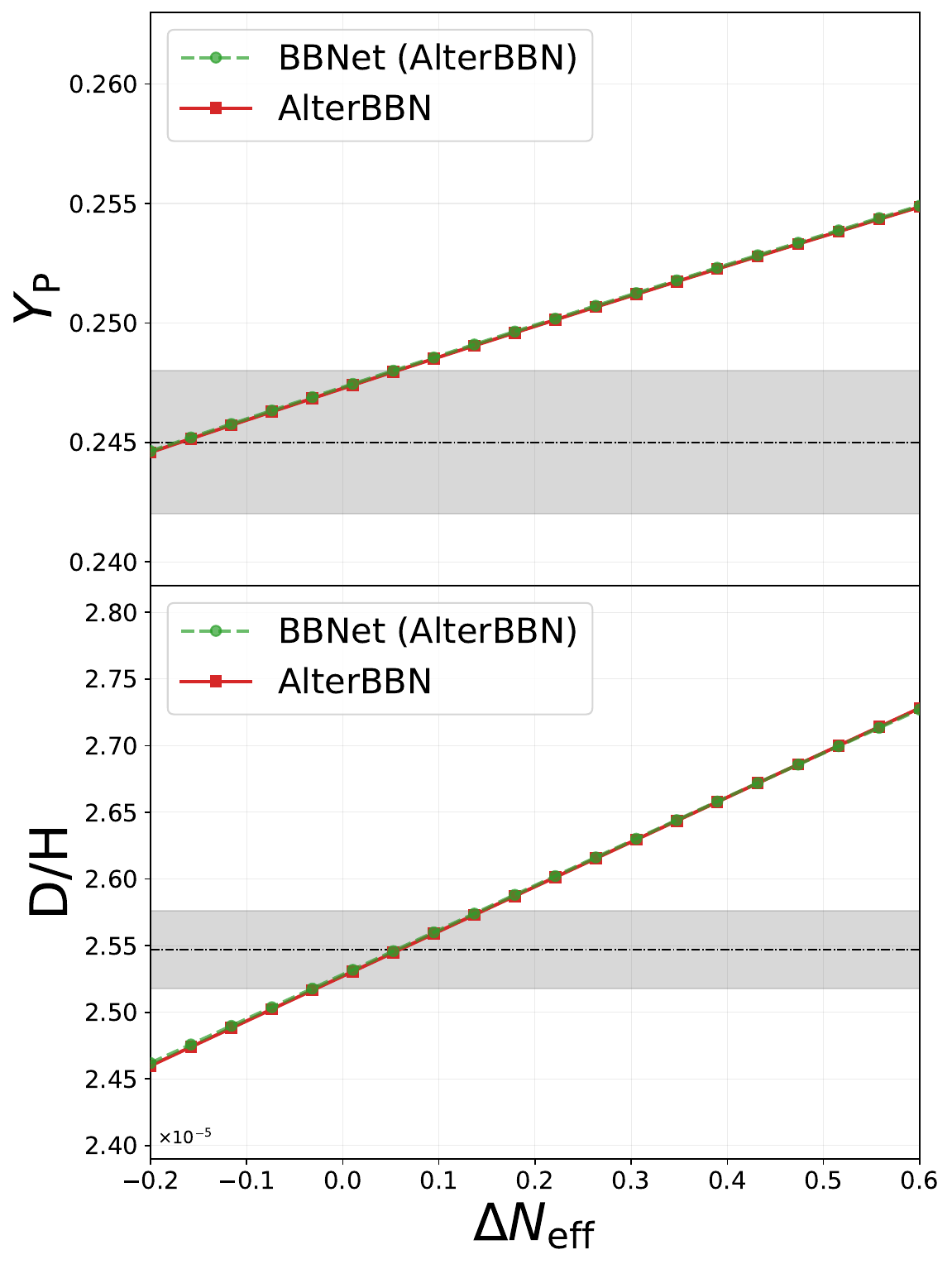}
    \end{minipage}
  }\hfill
  \subfloat[\label{fig:sens:omega_pe}]{%
    \begin{minipage}[t][\panelhB][c]{\panelwB}
      \centering
      \includegraphics[height=\panelhB,keepaspectratio]{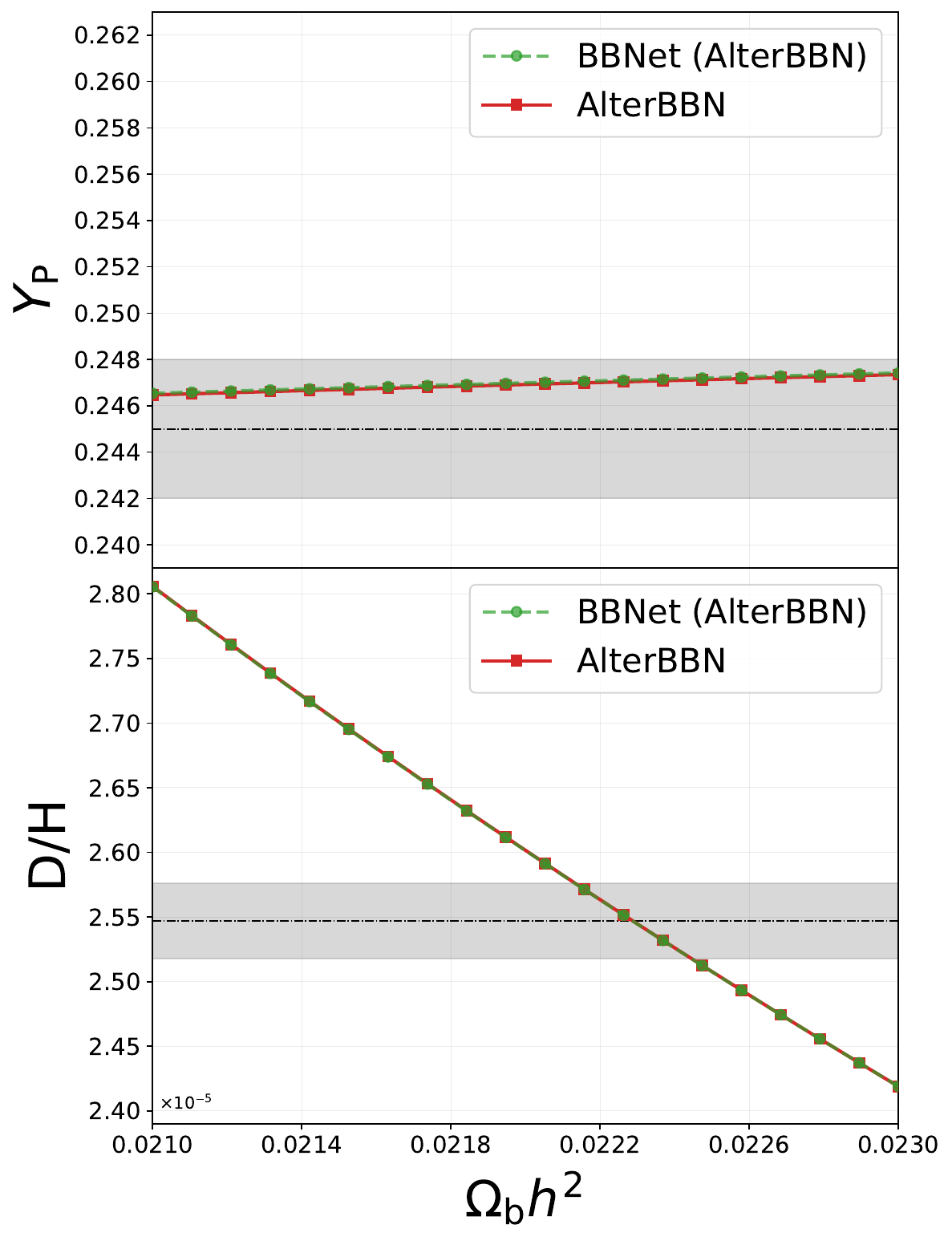}
    \end{minipage}
  }\hfill
  \subfloat[\label{fig:sens:tau_pe}]{%
    \begin{minipage}[t][\panelhB][c]{\panelwB}
      \centering
      \includegraphics[height=\panelhB,keepaspectratio]{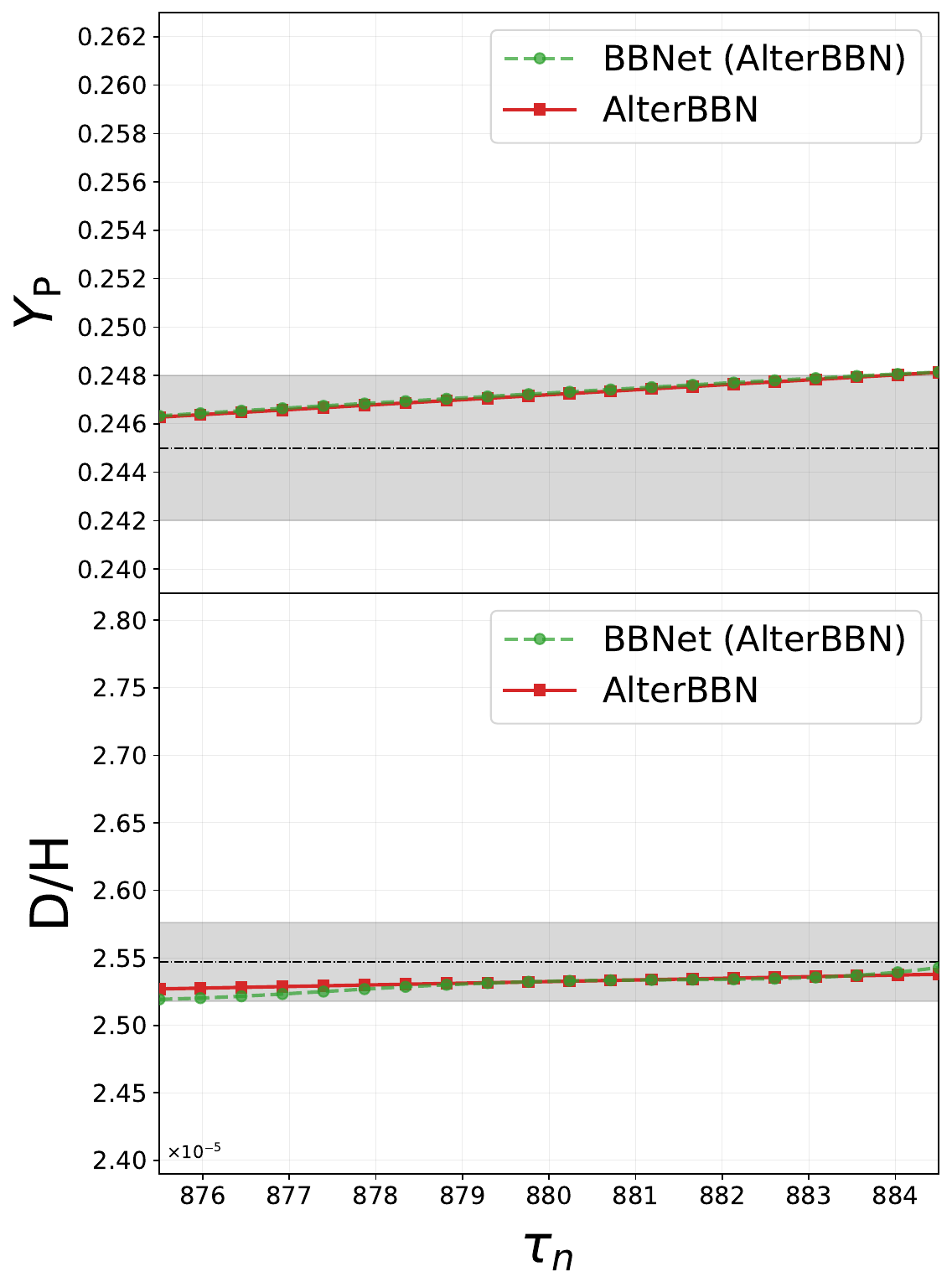}
    \end{minipage}
  }
  \caption{Parameter dependence of primordial abundances from AlterBBN and BBNet,
  similar to Fig.~\ref{fig:appendix_dependence_parthenope}.
  The top panels show the dependence of $Y_{\mathrm{P}}$
  and the bottom panels show the dependence of $\mathrm{D/H}$.
  Each panel varies a single input parameter
  while all remaining inputs are fixed to the fiducial values.}
  \label{fig:appendix_dependence_alterbbn}
\end{figure*}


\subsection{Dependence of primordial abundances on physical parameters}
\label{app:parameter_dependence}

Here we demonstrate the accuracy of \verb|BBNet| emulations from another perspective,
the dependence of $Y_\mathrm{P}$ and $\mathrm{D/H}$ on the physical parameters.
Using the test sets in both cases (\verb|PArthENoPE| and \verb|AlterBBN|),
we compare the predicted element abundances
with those obtained by the BBN calculations.
As illustrated in Figs.~\ref{fig:appendix_dependence_parthenope}
and \ref{fig:appendix_dependence_alterbbn}, each parameter is individually varied
while the others are fixed at their fiducial values:
$\kappa_{10}=0.0$, $\Delta N_\mathrm{eff}=0.0$,
$\Omega_\mathrm{b} h^2=0.02237$ \cite{2020A&A...641A...6P},
and $\tau_n=879.4\,\mathrm{s}$ \cite{Fields:2025}.
Our results show a perfect agreement
between the first-principles computations and the neural network emulations
over large variations in the values of parameters.

In this work, $\kappa_{10}$ and $\Delta N_\mathrm{eff}$
are the parameters that describe BSM physics.
Their relationships with the primordial element abundances
shown in Figs.~\ref{fig:appendix_dependence_parthenope}
and \ref{fig:appendix_dependence_alterbbn}
are essentially the same as those shown in Fig.~\ref{fig:bbn_code_comparison}.
In particular, $Y_\mathrm{P}$ increases monotonically with $\kappa_{10}$
and is highly sensitive to it, whereas the corresponding dependence
of $\mathrm{D/H}$ on $\kappa_{10}$ is comparatively weak.
By contrast, $\Delta N_\mathrm{eff}$ strongly influences both abundances
with a monotonically increasing trend.
As for the SM sector, changes in the baryon density, $\Omega_\mathrm{b} h^2$,
induce negligible changes in $Y_\mathrm{P}$
but a significant, monotonically decreasing effect on $\mathrm{D/H}$.
Variations in the neutron lifetime, however, have small impacts
on both $Y_\mathrm{P}$ and $\mathrm{D/H}$ with slightly increasing trends.



These results exhibit the robustness of \verb|BBNet|
in tasks that involve parameter sweeps, where rapid evaluations are essential. 
By retaining the outcomes of traditional numerical solvers
while offering a speed-up of several orders of magnitude,
\verb|BBNet| enables efficient, bias-free cosmological inferences with BBN data,
even when the prior ranges of parameters are wide.

\section{Comparison Between BBNet Emulations and Simplified BBN Calculations}
\label{sec:comparison}

Now that we have demonstrated the performance
of our neural network emulator \verb|BBNet|,
in this section we compare it with the traditional numerical approaches
applied in existing parameter inference efforts,
which must involve a trade-off between speed and accuracy.
In fact, to allow for MCMC analyses that converge in reasonable hours,
standard pipelines often rely on reduced nuclear networks
or simplified ODE integrators.
Although these approximations are able to meet the working level of speed,
they may introduce systematic biases that can propagate into final inference results.

The comparison is carried out in two folds.
In Sec.~\ref{sec:mcmc} we show that when applied to realistic MCMC sampling,
the computational efficiency of \verb|BBNet| exceeds
those of simplified BBN calculations by orders of magnitude.
It implies that the speed-accuracy trade-off utilized by simplified numerical solvers
becomes unnecessary when neural network emulators come online (Sec.~\ref{sec:tradeoff}).
Sec.~\ref{accuracy_vs} analyzes the error distributions
that result from the two approaches.
Systematic biases characteristic of simplified first-principles methods
are not found in the \verb|BBNet| results.

\subsection{Sampling efficiency in MCMC inferences}
\label{sec:mcmc}

To quantify the computational acceleration provided by \verb|BBNet|,
we first conduct a set of benchmark tests for our emulator
using MCMC sampling of a single chain with $1000$ steps.
Next, we perform realistic MCMC analyses based on the measured values of $Y_\mathrm{P}$ and D/H
reported by PDG 2024 \cite{Fields:2025}.

For comparison, we choose commonly used configurations
of \verb|PArthENoPE| and \verb|AlterBBN| as examples of simplified BBN calculations,
which reflect the speed-accuracy trade-off.
For \verb|PArthENoPE|, we compare \verb|BBNet| with its ``small'' network
that takes into account 9 nuclides and 40 reactions only \cite{2008CoPhC.178..956P}.
Recall that the corresponding \verb|BBNet| emulator reproduces
results of the benchmark ``complete'' network.
For \verb|AlterBBN|, we contrast \verb|BBNet|
with the standard second-order Runge-Kutta method implemented in the code
(setting  ``failsafe=3'' in \verb|AlterBBN|), 
dubbed ``RK2'' here \cite{2018arXiv180611095A}.
This RK2 mode is faster but less accurate than the benchmark RK2\_halfstep mode
to which we anchor the emulator.

\begin{table*}[htbp]
\caption{Comparison of computational time costs
between the numerical methods and BBNet in MCMC sampling tests.}
\label{tab:mcmc_speed_comparison}
\begin{ruledtabular}
\begin{tabular}{lcccc}
\textbf{Code} & \textbf{Steps} & \textbf{Chains} & \textbf{Total Runtime (s)} & \textbf{Speed-up} \\
\midrule
\multicolumn{5}{c}{\textit{Comparison Group 1: PArthENoPE}} \\
PArthENoPE (complete network) & 1000 & 1 & 16860.39 & $1\times$ \\
PArthENoPE (small network)  & 1000 & 1 & 830.38  & ~$ 20.3\times$ \\
BBNet Emulator (trained on \verb|PArthENoPE|) & 1000 & 1 & 2.29 & $7.4\times10^{3}$ \\
\midrule
\multicolumn{5}{c}{\textit{Comparison Group 2: AlterBBN}} \\
AlterBBN (RK2\_halfstep) & 1000 & 1 & 21267.83 & $1\times$ \\
AlterBBN (RK2) & 1000 & 1 & 569.49  & $37.3\times$ \\
BBNet Emulator (trained on \verb|AlterBBN|) & 1000 & 1 & 7.64 & $2.7\times10^{3}$ \\
\bottomrule
\end{tabular}
\end{ruledtabular}
\end{table*}

For the fixed-step MCMC sampling tests,
our results are summarized in Table~\ref{tab:mcmc_speed_comparison}.
It demonstrates that \verb|BBNet| provides the optimal solution
with its tremendous advantage in computational efficiency.
In this MCMC setting, the emulator achieves a speed-up of $\gtrsim10^{3}$
relative to the benchmark, full BBN methods
(the complete network of \verb|PArthENoPE|
and the RK2\_halfstep mode of \verb|AlterBBN|).
More importantly, it outperforms the simplified methods
by a factor of $\mathcal{O}(10^2)$.
These findings render the reduced numerical methods unnecessary:
\verb|BBNet| is superior to the latter in both efficiency and accuracy.


For realistic applications to cosmological inferences, 
we employ the following standard Gaussian log-likelihood
for the primordial abundance of each element: 
\begin{equation}
\ln \mathcal{L} = -\frac{1}{2}
\frac{\left[Y_{\mathrm{obs}}-Y_{\mathrm{th}}(\boldsymbol{\theta})\right]^2}
{\sigma_{\mathrm{obs}}^2+\sigma_{\mathrm{th}}^2}
\label{eq:abundance_like}
\end{equation}
up to a constant, where $\sigma_{\mathrm{obs}}$ denotes the observational uncertainty
and $\sigma_{\mathrm{th}}$ the theoretical uncertainty associated with the first-principles prediction of the primordial abundance,
here $\boldsymbol{\theta}=\{\Omega_\mathrm{b}h^2,\,\tau_n,\,\Delta N_\mathrm{eff},\,\kappa_{10}\}$.
Based on this extended cosmological model, 
we consider the following two scenarios: Case~A, full analyses with all four free parameters, 
and Case~B, conditional two-parameter analyses that only vary $\Delta N_\mathrm{eff}$ and $\kappa_{10}$, fixing $\Omega_\mathrm{b}h^2$ and $\tau_n$ to their fiducial values. 
We perform a series of MCMC inferences using the \verb|Cobaya| framework \cite{2019ascl.soft10019T,2021JCAP...05..057T}. 
We enforce the default Gelman-Rubin convergence criterion of $R-1<0.01$ \cite{1992StaSc...7..457G}.

\begin{figure*}[t]
\centering
\subfloat{
    \includegraphics[width=0.47\textwidth]{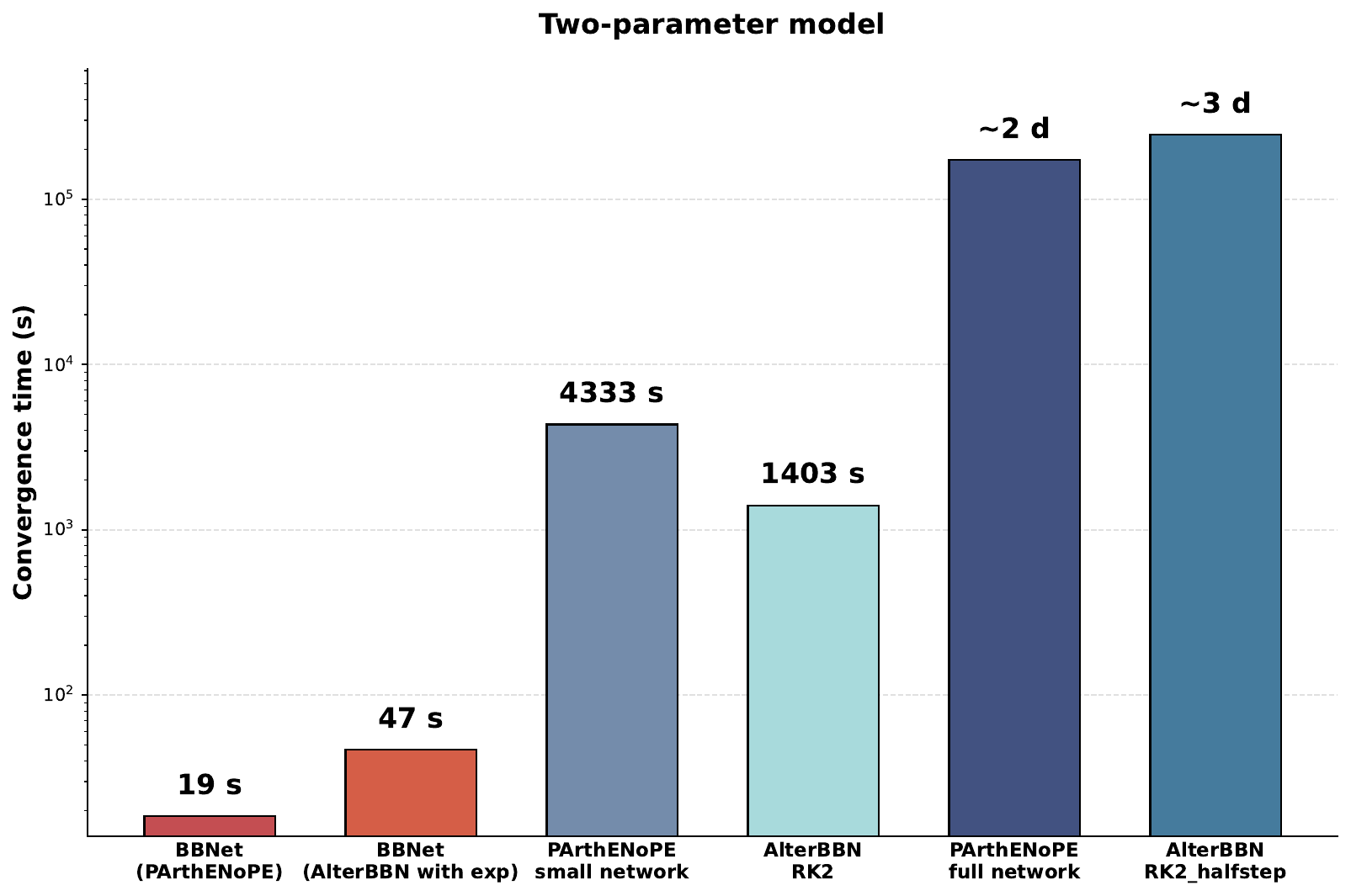}
    \label{fig:mcmc_convergence_2d}
}
\hfill
\subfloat{
    \includegraphics[width=0.47\textwidth]{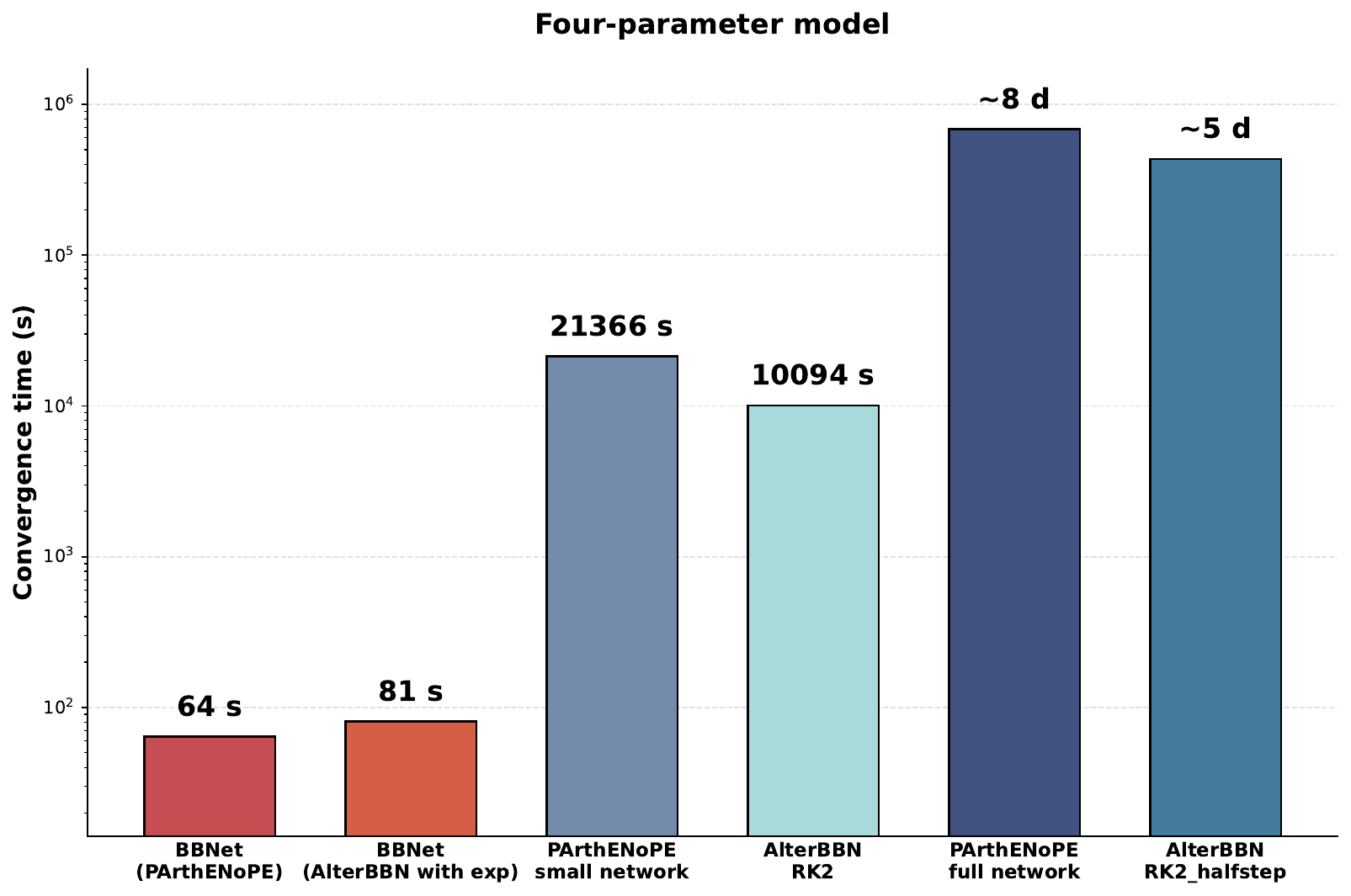}
    \label{fig:mcmc_convergence_4d}
}
\caption{Wall time required to reach convergence for Cobaya MCMC inferences. \emph{Left Panel}: conditional two-parameter model (Case~B) that only varies $\Delta N_\mathrm{eff}$ and $\kappa_{10}$. \emph{Right Panel}: full four-parameter cosmological model (Case~A). 
Both panels show the sheer contrast between the measured runtime with the BBNet emulators and those with the numerical BBN codes; both full and simplified configurations of PArthENoPE and AlterBBN are shown.
All MCMC chains are run with the same setting.}
\label{fig:mcmc_convergence}
\end{figure*}

The resulting convergence times are illustrated in Fig.~\ref{fig:mcmc_convergence}.
For the two-parameter model, \verb|BBNet| reaches convergence in only ${\sim}\,47$\,s, 
significantly faster than the analyses using even the simplified numerical methods, 
i.e., the small-network mode of \verb|PArthENoPE| and the RK2 mode of \verb|AlterBBN|. 
With the full BBN solvers, it takes as long as several days to converge.
For the four-parameter model, the MCMC runs with \verb|BBNet| still converge quickly (within ${\sim}\,81$\,s),
whereas even the simplified BBN solvers cost an unbearably long runtime of $\gtrsim 10^4$\,s.

As a result, the advance of \verb|BBNet| in computational efficiency will enable 
the multitude of exploratory cosmological inferences with different combinations of data sets and prior distributions,
a time-consuming or even impractical task for traditional numerical BBN solvers.

\begin{figure*}[t]
    \centering
    \subfloat[\label{fig:tradeoff_yp_main}]{
        \includegraphics[width=0.47\textwidth,keepaspectratio]{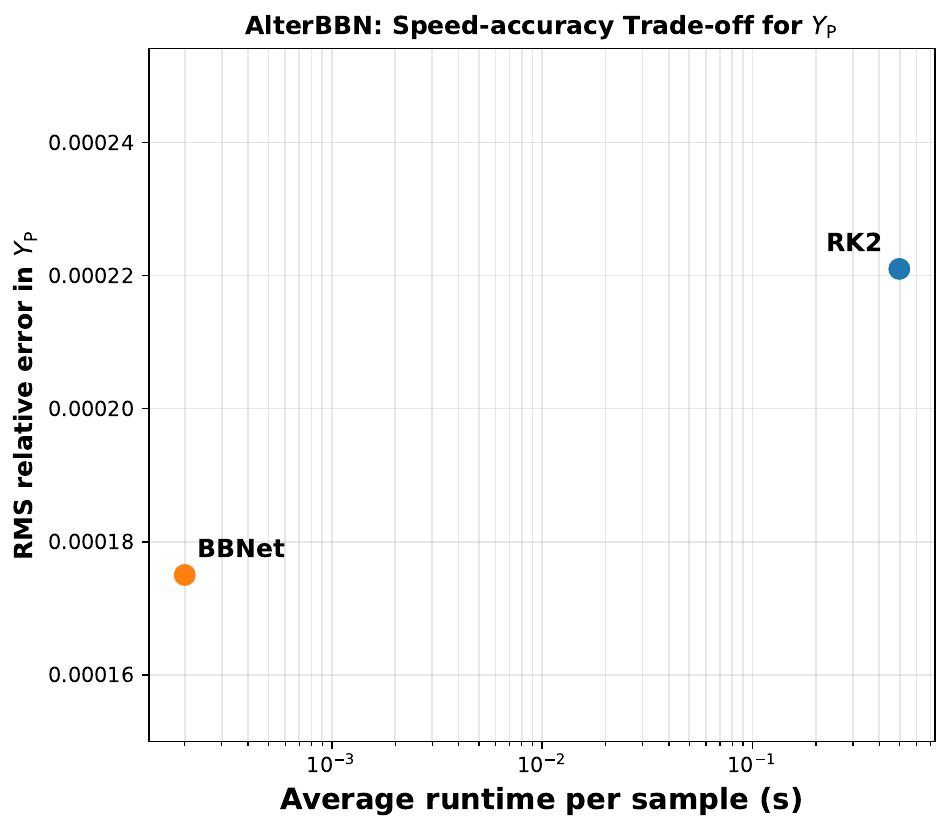}
    }\hfill
    \subfloat[\label{fig:tradeoff_dh_main}]{
        \includegraphics[width=0.47\textwidth,keepaspectratio]{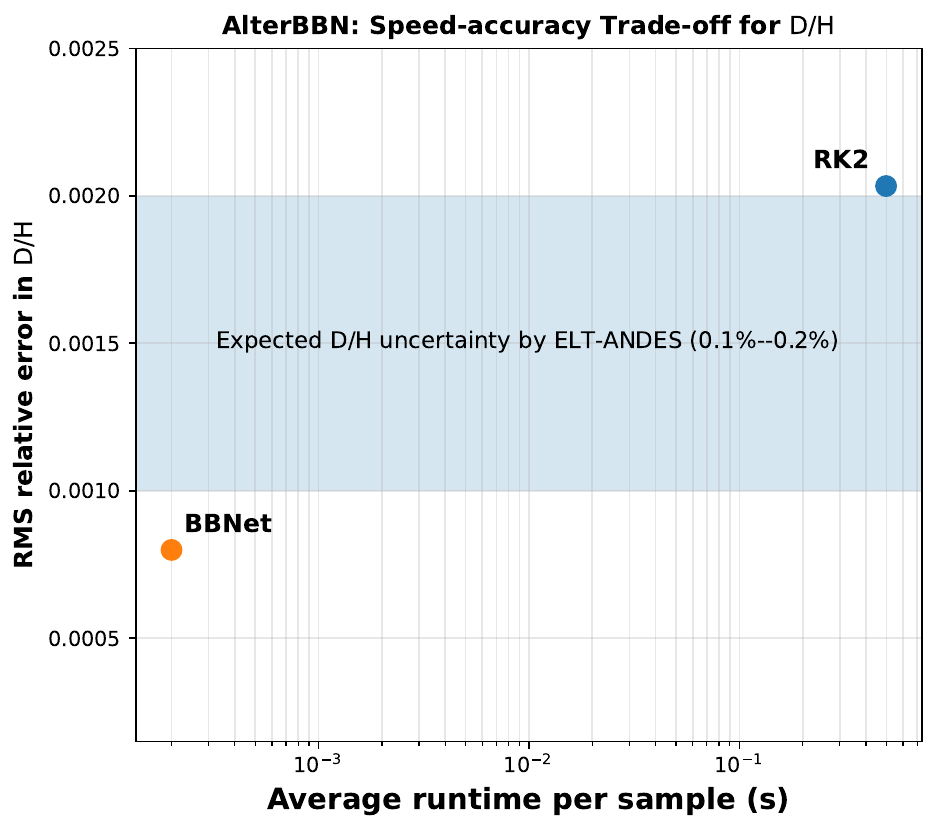}
    }
    \caption{Comparison of speed and accuracy between BBNet and the simplified RK2 mode of AlterBBN.
    The \emph{left panel} shows the RMS relative errors in $Y_\mathrm{P}$ and the average runtime of each method per sample.
    The \emph{right panel} shows the D/H case.
    In both panels, the relative errors are benchmarked against the high-accuracy AlterBBN RK2\_halfstep solver.
    The shaded horizontal band in the right panel denotes the expected $0.1\%$--$0.2\%$ uncertainty 
    of the future ELT-ANDES measurements of D/H \cite{2024ExA....57....5M}. 
    }
    \label{fig:error_hist_tradeoff}
\end{figure*}

\subsection{Speed-accuracy trade-off}
\label{sec:tradeoff}

Given the speed advantage of \verb|BBNet| in realistic parameter inferences,
if our neural network emulators also produce more accurate predictions than the simplified BBN solvers do,
the speed-accuracy trade-off embodied in the latter will no longer be needed.
Here we demonstrate this point on a per-sample basis.
Fig.~\ref{fig:error_hist_tradeoff} 
shows the RMS relative errors in the predicted primordial abundances
and the average wall time per evaluation
for both \verb|BBNet| and the \verb|AlterBBN| RK2 solver. 
The benchmark method that provides the ``true'' values
in both panels is the high-accuracy RK2\_halfstep solver of \verb|AlterBBN|.
For both $Y_\mathrm{P}$ and D/H, \verb|BBNet| leads in both efficiency and accuracy,
and its RMS errors are well below the current observational uncertainties.
In addition, as shown in the right panel of Fig.~\ref{fig:error_hist_tradeoff},
the emulation errors of D/H by \verb|BBNet| are still subdominant
compared with the expected measurement uncertainties from the future ELT-ANDES spectrograph \cite{2024ExA....57....5M},
while the simplified numerical solver produces larger errors.

The results of these diagnostics prove that \verb|BBNet| has the edge on both speed and accuracy. 
Hence, neural network emulation can well replace the traditional practice
of using simplified solvers that rely on the speed-accuracy trade-off.

\begin{table}[ht]
  \centering
  \small
  \setlength{\tabcolsep}{4pt}
  \caption{Relative percentage error metrics (mean, median, RMSPE)
  for the comparison between the numerical methods and the emulator,
  grouped by the data sets (PArthENoPE and AlterBBN).}
  \label{tab:error_summary_combined}
  \begin{tabular}{llrrr}
    \toprule
    Abundance & Models & \multicolumn{3}{c}{Percentage Error (\%)} \\
    \cmidrule(lr){3-5}
              &        & Mean & Median & RMS \\
    \midrule
    \multicolumn{5}{c}{\emph{PArthENoPE}} \\
    \addlinespace[2pt]
    \multirow{2}{*}{$Y_\mathrm{P}$}
      & small network  & $ -0.0083$ & $ -0.0083$ & $0.0088$ \\
      & BBNet      & $ -0.0011$ & $ -0.0005$ & $0.0158$ \\
    \multirow{2}{*}{$\mathrm{D/H}$}
      & small network  & $ 0.0768$ & $ 0.0765$ & $0.0771$ \\
      & BBNet      & $ 0.0013$ & $ -0.0007$ & $0.0503$ \\
    \addlinespace[4pt]
    \multicolumn{5}{c}{\emph{AlterBBN}} \\
    \addlinespace[2pt]
    \multirow{2}{*}{$Y_\mathrm{P}$}
      & RK2        & $-0.0109$ & $-0.0089$ & $0.0221$ \\
      & BBNet      & $-0.0016$ & $-0.0011$ & $0.0175$ \\
    \multirow{2}{*}{$\mathrm{D/H}$}
      & RK2        & $-0.1563$ & $-0.1368$ & $0.2033$ \\
      & BBNet      & $0.0002$ & $0.0014$ & $0.0799$ \\
    \bottomrule
  \end{tabular}
\end{table}

\subsection{Residual error distribution}
\label{accuracy_vs}

Finally, we draw a further comparison between \verb|BBNet| and the simplified solvers
with respect to the accuracy argument.
As mentioned above, the speed-accuracy trade-off
exploited by the latter
often leads to systematic biases in the predicted abundances.
In contrast, a robust emulator such as \verb|BBNet| is designed
to reproduce the ground truth of the full numerical calculations
without introducing systematic shifts in residual errors.
Using the same high-accuracy benchmarks,
we analyze the distributions of residual errors
in the results of both \verb|BBNet| emulations and simplified calculations
(the small network of \verb|PArthENoPE| and the RK2 mode of \verb|AlterBBN|).
We evaluate the same metrics as in Sec.~\ref{sec:accuracy_comparison}, e.g., MPE and RMSPE.
The results are summarized in Table~\ref{tab:error_summary_combined}.

\begin{figure*}[htpb]
  \centering
  \subfloat[\label{fig:error_yp}]{%
    \includegraphics[width=0.48\textwidth,keepaspectratio]{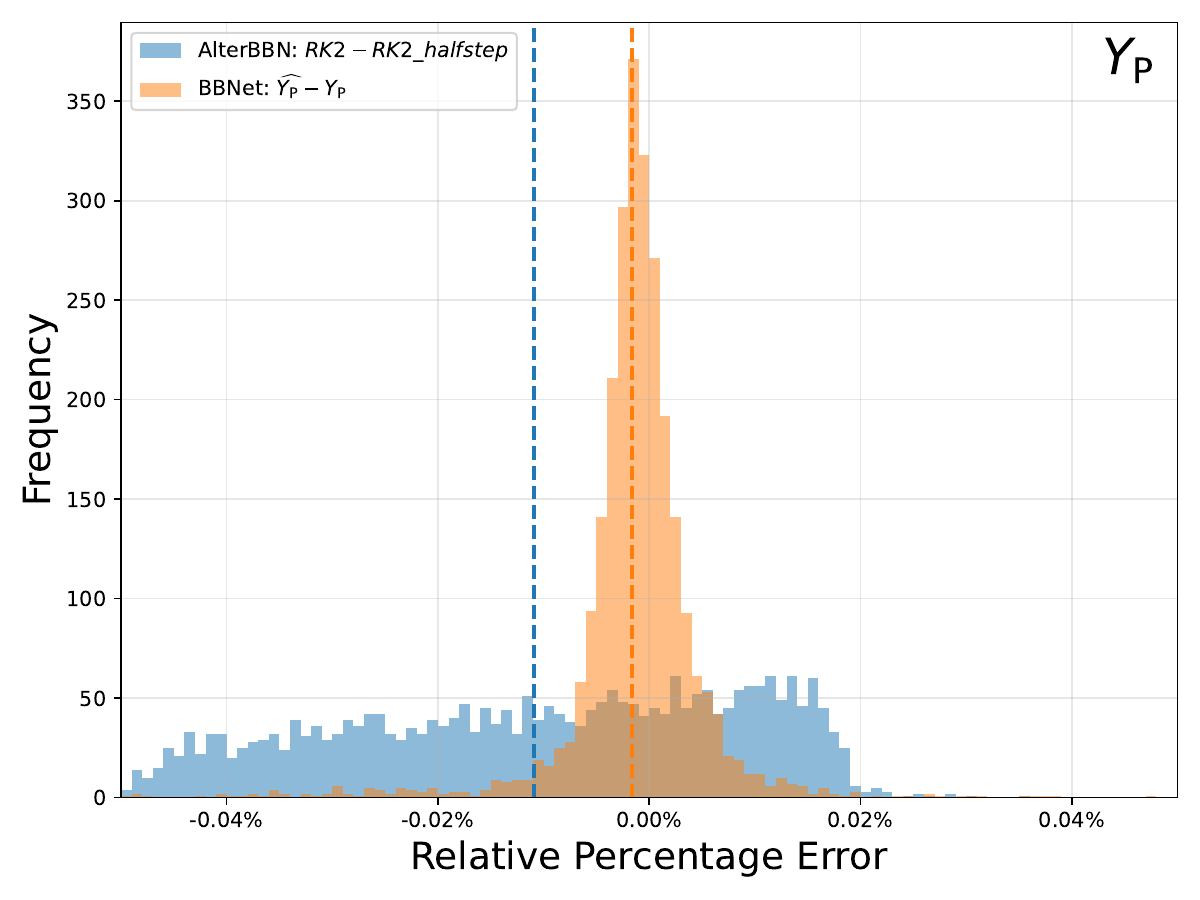}
  }\hfill
  \subfloat[\label{fig:error_dh}]{%
    \includegraphics[width=0.48\textwidth,keepaspectratio]{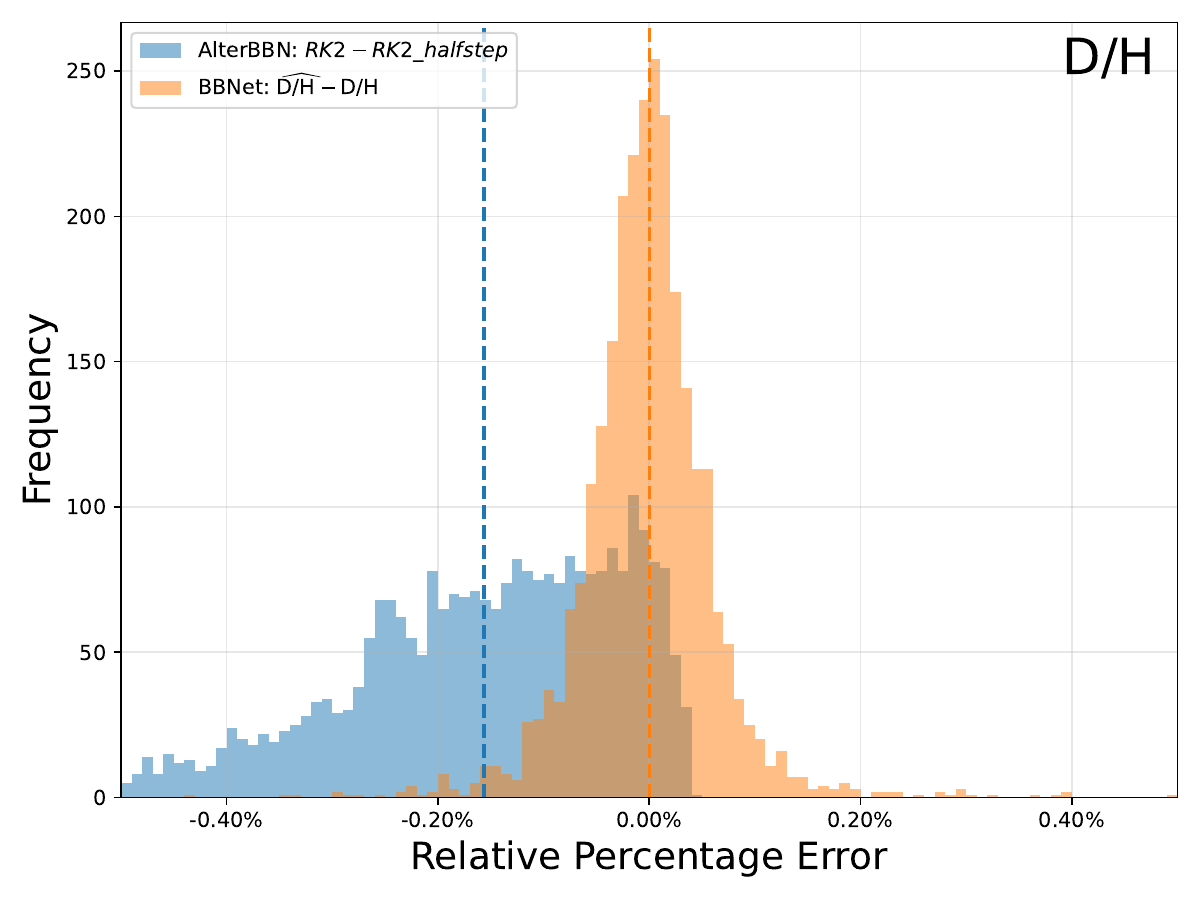}
  }
    \caption{Relative percentage error distributions in the AlterBBN test set.
    The left panel shows the errors for $Y_{\mathrm{P}}$, and the right panel shows the errors for $\mathrm{D/H}$. The errors are computed relative to the high-accuracy reference results obtained with the RK2\_halfstep mode. Blue histograms show the residuals between the RK2 mode and the standard RK2\_halfstep reference. Orange histograms show the residuals between the BBNet predictions and the RK2\_halfstep reference, quantifying the emulator’s interpolation error. Dashed vertical lines in each color indicate the corresponding mean signed error. The horizontal axes are restricted to $\pm 0.05\%$ for $Y_{\mathrm{P}}$ and $\pm 0.5\%$ for $\mathrm{D/H}$ to resolve the fine structure of the distributions.
    }
    \label{fig:error_comparison}
\end{figure*}


For the \verb|PArthENoPE| data set
(see the upper block of Table~\ref{tab:error_summary_combined}),
\verb|BBNet| shows a distinct advantage
in eliminating systematic biases (described by MPE).
The small network approximation introduces an average offset of $\sim 0.08\%$
for $\mathrm{D/H}$, a deviation that is consistent throughout the parameter space,
while \verb|BBNet| reduces the MPE to a negligible level ($0.0013\%$).

The superiority of the emulator is further demonstrated
by the comparison within the \verb|AlterBBN| data set,
as shown in the lower block of Table~\ref{tab:error_summary_combined}
and Fig.~\ref{fig:error_comparison}.
The faster but less accurate RK2 solver exhibits a considerable systematic drift,
underestimating both $Y_\mathrm{P}$ and $\mathrm{D/H}$.
The error distributions of the \verb|BBNet| predictions,
conversely, are centered around zero (with MPEs~$\lesssim10^{-3}\,\%$),
as illustrated in Fig.~\ref{fig:error_comparison}.
Thus, our emulator faithfully reproduces
the solutions of the benchmark RK2\_halfstep method
without inducing any systematic skew as does the RK2 integrator.
In addition, \verb|BBNet| also significantly reduces
scatters in the error distributions,
e.g., lowering the RMSPE in the $\mathrm{D/H}$ results
by a factor of $\sim2.5$ relative to the RK2 solver.
The emulator achieves an RMSPE of $\sim10^{-2}\,\%$
for both $Y_\mathrm{P}$ and $\mathrm{D/H}$,
two orders of magnitude smaller than the relative errors
from astronomical measurements; cf.~Eq.~(\ref{eq:obs_constraints}).
Therefore, when applied to a likelihood analysis,
\verb|BBNet| can yield correct constraints on the physical parameters
that are determined only by observational data
without introducing any computational errors.

These results confirm that \verb|BBNet| can supersede
simplified numerical schemes to generate primordial element abundances.
By efficiently reproducing the solutions of full BBN calculations
with negligible systematic biases and scatter,
the emulator serves as a promising tool for cosmological inferences.


\section{Conclusion}
\label{sec:conclusion}

We have presented \verb|BBNet|, a deep learning emulator designed
to accelerate BBN calculations without compromising physical accuracy.
The network employs a residual multi-head attention architecture
to capture the nonlinear dependencies of primordial abundances on physical parameters.
We trained the model using Latin-hypercube samples
derived from two independent BBN codes, \verb|PArthENoPE| and \verb|AlterBBN|,
ensuring consistent treatments of nuclear physics between the two data sets.
The physical parameters include those from the Standard Model
as well as those that describe BSM physics, i.e., dark radiation and a stiff fluid.

Our results demonstrated that \verb|BBNet| can effectively resolve
the traditional trade-off between computational speed and theoretical accuracy.
For both BBN codes, the emulator successfully reproduced the benchmark predictions
of the full numerical methods with negligible biases and scatter.
Particularly, \verb|BBNet| eliminated the systematic offsets
inherent in simplified numerical calculations
(such as the small network of \verb|PArthENoPE|). 
As measurement precision of the primordial element abundances
improves with future facilities such as ELT-ANDES, 
accurate theoretical predictions with subdominant contributions to the total error budget
will become crucial to cosmological inferences. The extraordinary emulation accuracy of
\verb|BBNet| (of $<0.1\,\%$ RMSPE) makes it well prepared for that regime.
In terms of efficiency, the emulator achieves a speed-up of $\mathcal{O}(10^2)$
relative to the simplified methods
and up to $\mathcal{O}(10^{4})$ compared with the full benchmark solvers.
Based on these advantages, we advocate for the replacement
of traditional approximate approaches
by fast, high-fidelity neural network emulation.

Indeed, the combination of accuracy and efficiency held by \verb|BBNet|
has removed the primary computational bottleneck in large-scale parameter inferences.
\verb|BBNet| provides an optimal solution
to the forward modeling of primordial element abundances within Bayesian pipelines,
enabling rigorous explorations of high-dimensional parameter spaces.
Furthermore, our framework remains flexible:
Unlike static interpolation tables, the network can be efficiently retrained
to accommodate updated nuclear reaction rates or expanded prior ranges.

The \verb|BBNet| framework presented here can be generalized
to more detailed BBN studies,
such as those involving nuclear reaction rates as free parameters
or other BSM physics.
Fast and accurate emulation of primordial element abundances
will help shed light on the complex relationship
between the BBN observables and any relevant physics,
e.g., the \emph{real} theoretical uncertainties
due to poorly-determined nuclear rates (rather than computational errors).
Meanwhile, we also plan to extend the emulator
to predict the primordial abundances of helium-3 and lithium-7,
which will facilitate a reexamination of the famous lithium problem.

All source code, pre-trained models, and data sets used in this work
are publicly available at the link in Ref.~\cite{BBNet_repo}.

\section{ACKNOWLEDGMENTS}

F.Z. gratefully acknowledges the continuous support and valuable advice of the MIT LIGO Laboratory. This work was supported in part by the National Natural Science Foundation of China (Grant No. 62372409) and the Ministry of Science and Technology of the People’s Republic of China (Grant No. 2023ZD0120704 under Project No. 2023ZD0120700). B.L. is supported by the Guangxi Natural Science Foundation (Grant No.~2023GXNSFBA026114), Guangxi Key R\&D Program (Guangxi Funeng Action Plan, Grant No.~Guike FN2504240040), the Guangxi Science and Technology Innovation Platform Program (Leitai Action Plan, Grant No.~Guike LT2600640026) and the National Natural Science Foundation of China (Grants No.~12203012, No.~12494575). This work is also supported by the ``Guangxi Highland of Innovation Talents'' Program.

\section{Data Availability}
The data that support the findings of this article are openly available \cite{BBNet_repo}.

\appendix
\section{Hierarchical Expert Mode for Adaptive Refinement}
\label{experts}

To address the wide dynamic range and complex distribution
of the \verb|AlterBBN| data set,
we implement a two-stage hierarchical inference strategy.
The architecture consists of a global \emph{base} model, trained on the complete dataset, augmented by two domain-specialized \emph{expert} models trained on filtered subsets of the parameter space. All three networks share an identical backbone architecture (\verb|ResMLPWithAttn|).
Each model checkpoint is stored alongside its corresponding normalization statistics (scalers) to ensure that the input-output transformations remain strictly reversible and consistent with the specific training distribution of each expert.

During the inference phase, the framework employs a hard-gating mechanism (activated via the \verb|--exp| flag). The base model first generates a preliminary prediction, $\widehat{\mathrm{D/H}}_{\text{base}}$. This value acts as a decision variable to route the input to the appropriate specialist network or to accept the base prediction, according to the following rule:
\begin{equation}
  \widehat{\mathrm{D/H}}_\mathrm{out} = 
  \begin{cases} 
    \text{Expert 2} & \text{if } \widehat{\mathrm{D/H}}_{\text{base}} \in [10^{-7},\,10^{-5}), \\
    \text{Expert 1} & \text{if } \widehat{\mathrm{D/H}}_{\text{base}} \in [10^{-5},\,10^{-3}], 
  \end{cases}
\end{equation}
For routed samples, the raw inputs are renormalized using the target expert's specific statistics before being processed by the expert network.
This band-specific renormalization is essential because the statistical moments of the filtered subsets differ significantly from the global distribution, and using global scalers for local experts would degrade precision.

We quantify the computational overhead of this approach in terms of per-sample wall-clock time and theoretical floating-point operations (FLOPs).
In the \verb|AlterBBN| test run ($N=3000$),
the routing fraction is close to unity ($f\!\approx\!1$),
which means that nearly all samples trigger a second forward pass.
The resulting resource consumption is summarized in Table~\ref{tab:exp_cost_numeric}.
Since the expert mechanism requires loading up to three distinct sets of weights,
the memory cost increases by a factor of $\sim3$ compared to the base-only mode,
although this can be mitigated to $\sim2$ times via lazy loading.

\begin{table}[ht]
  \caption{Comparison of computational resources and latency
  between the \emph{base-only} mode and the \emph{expert} mode.
    Wall times are measured per sample over repeated trials.
    FLOPs are calculated for a single inference pass given the specific backbone configuration.
    The memory column reports the FP32 model weight footprint plus the storage required for the normalization scalers.}
  \label{tab:exp_cost_numeric}
  \begin{ruledtabular}
  \begin{tabular}{lccr}
    \textbf{Inference Mode} & \textbf{Latency ($\mu$s)} & \textbf{FLOPs} & \textbf{Memory} \\
    \hline
    Base only
      & 36.5--56.2
      & $4.03\times10^{8}$
      & $\ 768$\,MiB\\
    Expert mode
      & 74.6--97.3
      & $8.05\times10^{8}$
      & $\ 2.25$\,GiB \\
  \end{tabular}
  \end{ruledtabular}
\end{table}

\bibliography{refs_new}
\end{document}